\documentclass[aps,pra,twocolumn,superscriptaddress]{revtex4-1}

\usepackage[english]{babel}
\usepackage{amsmath}
\usepackage{amsthm}
\usepackage{amssymb}
\usepackage{mathtools}
\usepackage{ragged2e}
\usepackage{etoolbox}
\usepackage{booktabs}
\usepackage{graphicx}
\usepackage{xcolor}
\definecolor{xblue}{HTML}{3A56A5}
\definecolor{cerulean}{HTML}{08457E}
\definecolor{xgolden}{HTML}{D39E2A}
\usepackage{color}
\usepackage{hyperref}
\usepackage{cases}
\usepackage{caption}
\usepackage{enumerate}
\usepackage{enumitem}
\usepackage{multirow}
\hypersetup{
  colorlinks   = true,  
  urlcolor     = blue,  
  linkcolor    = blue,  
  citecolor    = red    
}

\usepackage[font=small,labelfont=bf,justification=justified,format=plain]{caption}
\usepackage{natbib}
\usepackage{pgfplots}
\pgfplotsset{compat=1.18}
\usepackage{subcaption}
\usepackage{url}
\newcommand{\ket}[1]{ | #1 \rangle }
\newcommand{\bra}[1]{ \langle #1 | }

\DeclareMathOperator*{\argmax}{arg\,max}

\theoremstyle{definition}

\theoremstyle{remark}

\usepackage{scalerel,stackengine}
\stackMath
\newcommand\reallywidehat[1]{%
\savestack{\tmpbox}{\stretchto{%
  \scaleto{%
    \scalerel*[\widthof{\ensuremath{#1}}]{\kern-.6pt\bigwedge\kern-.6pt}%
    {\rule[-\textheight/2]{1ex}{\textheight}}
  }{\textheight}%
}{0.5ex}}%
\stackon[1pt]{#1}{\tmpbox}%
}
\usepackage{comment}

\begin{document}

\title{Geometrical Approach to Logical Qubit Fidelities of Neutral Atom CSS Codes}

\author{J.J. \surname{Postema}}
\thanks{Corresponding author} \email{j.j.postema@tue.nl}
\affiliation{Department of Applied Physics and Science Education, Eindhoven University of Technology, P.~O.~Box 513, 5600 MB Eindhoven, The Netherlands}
\affiliation{Eindhoven Hendrik Casimir Institute, Eindhoven University of Technology, P.~O.~Box 513, 5600 MB Eindhoven, The Netherlands}

\author{S.J.J.M.F. \surname{Kokkelmans}}
\affiliation{Department of Applied Physics and Science Education, Eindhoven University of Technology, P.~O.~Box 513, 5600 MB Eindhoven, The Netherlands}
\affiliation{Eindhoven Hendrik Casimir Institute, Eindhoven University of Technology, P.~O.~Box 513, 5600 MB Eindhoven, The Netherlands}

\date{\today}

\begin{abstract}
Encoding quantum information in a quantum error correction (QEC) code enhances protection against errors. Imperfection of quantum devices due to decoherence effects will limit the fidelity of quantum gate operations. In particular, neutral atom quantum computers will suffer from correlated errors because of the finite lifetime of the Rydberg states that facilitate entanglement. Predicting the impact of such errors on the performance of topological QEC codes is important in understanding and characterising the fidelity limitations of a real quantum device. Mapping a QEC code to a $\mathbb{Z}_2$ lattice gauge theory with disorder allows us to use Monte Carlo techniques to calculate upper bounds on error rates without resorting to an optimal decoder. In this Article, we adopt this statistical mapping to predict error rate thresholds for neutral atom architecture, assuming radiative decay to the computational basis, leakage and atom loss as the sole error sources. We quantify this error rate threshold $p_\text{th}$ and bounds on experimental constraints, given any set of experimental parameters.
\end{abstract}

\maketitle

\section{Introduction}

Quantum computing is currently in the \textit{Noisy Intermediate Scale Quantum} (NISQ) era, in which noise imposes a limit on the fidelity of qubits and gate operations \cite{nisq}. In order to build more robust qubits, quantum error correction has to be invoked \cite{terhal}. Mapping an ensemble of physical noisy qubits to a logical qubit enhances protection against errors. Surface codes are a class of topological \textit{quantum error correction} (QEC) codes that have been extensively studied \cite{Kitaevtoriccode, fowler, agameofsurfacecodes}. Only recently has it been demonstrated experimentally for the very first time, that QEC using surface codes can suppress logical error rates on near-term quantum devices \cite{googleAI, googleAI2}.\\

A promising candidate for quantum computers is an array of \textit{neutral atoms} trapped in optical tweezers. The quantum information is stored in long-lived atomic hyperfine clock states \cite{clock_kaufman, clock_thompson}, while entanglement is mediated through excitation to a high-$n$ Rydberg state. This system has several attractive features, such as identical qubits, long coherence times and a flexible geometry \cite{eiffel}. Recently, it has been shown that neutral atoms can be shuttled around using \textit{movable tweezers} with excellent preservation of coherence \cite{movabletweezers}, and error suppression with the Steane and toric code have been demonstrated using this technique \cite{lukinQEC}.\\

Statistical physics provides a powerful tool to analyse the performance of error correction codes \cite{spinglass}. It has been shown that there exists a duality between quantum codes and statistical mechanics, called the \textit{statistical mechanical mapping} \cite{topologicalquantummemory}. Mapping quantum codes to a random $\mathbb{Z}_2$ lattice gauge model reveals that the quantum error correction error rate threshold manifests itself as a \textit{second order phase transition}. In particular, the error probability is an order parameter that distinguishes two phases: an 'ordered' phase in which scaling up the code distance $d$ allows qubits to be driven towards arbitrarily low logical error rates, and a 'disordered' phase where errors irrevocably corrupt quantum information. This mapping requires no optimal decoder to evaluate the threshold.\\

Entanglement is a crucial aspect of fault-tolerant computation, but its consequences for error propagation are often not considered in phenomological error models such as a depolarisation channel. Multi-qubit gates are pipelines for crosstalk between qubits, but are also themselves inherent sources of \textit{correlated errors} \cite{saffman}. On a neutral atom quantum device, the instability of the Rydberg state that mediates entanglement, leakage outside of the computational basis and atom loss will be dominant sources of errors \cite{biasederasure}. Because they also facilitate measurement errors, their effect on the QEC code performance as the number of cycles increases, is significant.\\

In this Article, we provide a thorough analysis of the effect of entanglement errors on the performance of topological quantum error correction codes, tailored towards neutral atom quantum computers. Sec. \ref{sec:atoms} introduces the physics of neutral atoms, and showcases error correction protocols on neutral atom hardware. In Sec. \ref{sec:statmech},  we adopt  the statistical model of errors, and how mapping is achieved given a correlated error probability distribution. Results are given in Sec. \ref{sec:codeperformance}, where a Monte Carlo simulation of noisy neutral atoms is compared to a second order phase transition in a statistical mechanical model. In Sec. \ref{sec:summary}, we provide a summary of our work.

\section{Neutral atom error correction}\label{sec:atoms}

\begin{figure}[t!]
    \centering{
    \includegraphics[width=0.4\textwidth]{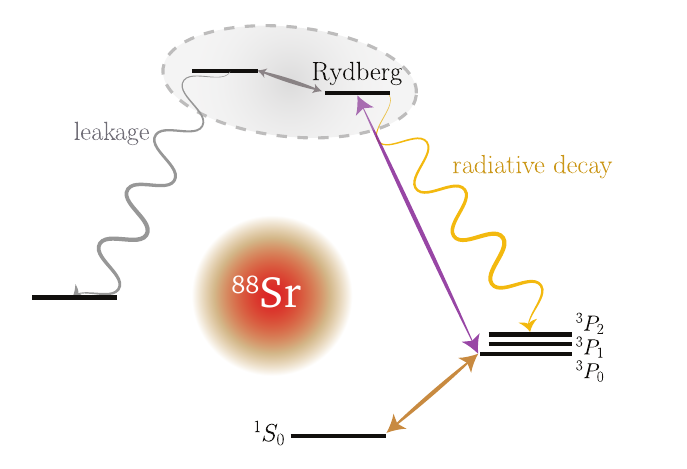}
    \captionsetup{justification=Justified}
    \caption{The $^{88}$Sr level scheme, with the computational basis states encoded in the clock states ($\ket{0} = \ket{^{1}S_0}$ and $\ket{1} = \ket{^{3}P_0}$). Two important decoherence processes have been highlighted: radiative decay to the computational manifold, and leakage outside the qutrit state submanifold. Unlabeled states depend on the choice of Rydberg state.}
    \label{fig:level}
    }
\end{figure}

Neutral atoms, such as $^{85}$Rb and $^{88}$Sr, often encode the qubit computational basis states in the electronic ground state manifold or clock states. These states have long coherence times on the order of seconds \cite{clock_kaufman, clock_thompson} and the clock transition has a narrow linewidth \cite{linewidth}. In the rest of this Article, we work in the $\{\ket{0}, \ket{1}, \ket{r}\}$ qutrit manifold of the $^{88}$Sr atom, which is endowed with the auxiliary Rydberg state $\ket{r}$ that facilitates long-range entanglement, which we leave unspecified. Fig. \ref{fig:level} shows the level scheme of $^{88}$Sr, in whose electronic states the $\ket{0}$- and $\ket{1}$-states have been embedded. The transition from $\ket{1}$ to $\ket{r}$ is driven by a single-photon process. Though we focus on ${}^{88}\text{Sr}$, our method applies to all neutral atoms with a similar level structure.\\

The architecture of a neutral atom quantum computer provides a scalable platform to implement quantum error correction codes. Topological codes such as the surface code, the toric code and various colour codes are favourable candidates for experimental proof-of-principle demonstrations for quantum error correction \cite{googleAI, lukinQEC}. Movable tweezers have been shown to transport atoms without significant loss of fidelity \cite{movabletweezers}, enabling favourable properties such as having dedicated read-out zones for local measurement \cite{readout}, and enabling flexible long-range entanglement with applications to for example low-density parity check codes \cite{lukinLDPC, ibmLDPC}. A great limiting factor to the fidelity of neutral atom qubits is decoherence, driven by stochastic processes such as radiative decay and leakage, highlighted in Fig. \ref{fig:level}. The remainder of this Section is dedicated to modelling these processes.\\

\subsection{Error modelling}

Often, a phenomenological depolarisation model is employed to model errors, characterised by a single error parameter $p\in[0,1]$. Such an $n$-qubit depolarisation channel $\mathcal{L}: \mathbb{C}^{2^n\times 2^n}\mapsto\mathbb{C}^{2^n\times 2^n}$ is given by:
\begin{equation}\label{eq:depolar}
    \mathcal{L}(\rho) = (1-p)\rho + \frac{p}{3}\sum_i \sigma^i \rho \sigma ^i
\end{equation}
for arbitrary density matrices $\rho$ and a Pauli basis $\{\sigma^i\}_i = \{X, Y, Z\}$. Such a model is sufficient for many proof-of-principle simulations, but leaves out the intricate details of entanglement errors: correlated errors that may be spacetimelike separated in the circuit. The dominant source of these errors is the erroneous implementation of controlled logic (CZ) during the stabilisation process. Besides being conduits of errors, multi-qubit gates can also introduce new errors themselves, subsequently affecting other rounds of stabilisation. The order of stabilisation, and if all qubits are entangled simultaneously or not, will affect this correlation propagation, and will play an important role in the transpilation of quantum circuits in the fault-tolerant era.\\

Given is a noiseless unitary evolution $\mathcal{F}(\rho)=U\rho U^\dagger$. Introducing a noise channel with multi-qudit Kraus representatives $\{\mathfrak{D}_\mu\}$, we obtain a \textcolor{black}{decoherent} error channel
\begin{equation}
    \mathcal{E}(\rho) = \sum_{\mu\nu} \lambda_{\mu\nu} \mathfrak{D}_\mu\mathcal{F}(\rho)\mathfrak{D}_{\nu}^{\dagger}
\end{equation}
relative to the perfect coherent channel $\mathcal{F}(\rho)$, for some coefficients $[\lambda_{\mu\nu}]$. These Kraus operators \textcolor{black}{describe decoherent noise channels} and satisfy the \textcolor{black}{completeness relation}
\begin{equation}
    \sum_\mu \mathfrak{D}_\mu^\dagger \mathfrak{D}_\mu = \mathbb{I}_{m \;\text{qudits}}
\end{equation}
on the Hilbert space $\mathfrak{H}^m$ of $m$ qudits. The operators themselves are generated by single-qudit Kraus operators $E_{q_i}^{i}$ acting on qubit $i$, with index $q_i$. \textcolor{black}{Through a change of basis, we can transform this channel $\mathcal{E}(\rho)$ to the Pauli basis as}
\begin{equation}\label{eq:chi}
    \mathcal{E}(\rho) = \sum_{\mu\nu} \chi_{\mu\nu} \mathcal{P}_\mu\mathcal{F}(\rho)\mathcal{P}_{\nu}^{\dagger},
\end{equation}
and use the Pauli twirling approximation \cite{twirling, twirling2, sventwirl} \textcolor{black}{to obtain a channel which is diagonal in the $\{\mathcal{P}_\mu\}$-basis, by only keeping diagonal terms:}
\begin{equation}\label{eq:twirl}
    \mathcal{E}^{\text{twirl}}(\rho) = \sum_{\mu} \chi_{\mu\mu}\mathcal{P}_\mu \mathcal{F}(\rho) \mathcal{P}_\mu^\dagger.
\end{equation}
The validity of twirling is discussed in Appendix \ref{sec:twirl}.\\

The first parameter our model considers is radiative decay from the Rydberg state to $\ket{1}$, with decay width $\gamma$. While the blockade strength $V$ is another source of infidelity, it is not of any significant influence to our results since we assume $V/\Omega \gg 1$. For a decay from the Rydberg state to the $\ket{1}$-state, the Kraus operators read
\begin{equation}\label{eq:kraus}
    E_1 = \begin{bmatrix}
1 & 0 & 0\\
0 & 1 & 0\\
0 & 0 & \sqrt{1-\gamma \delta t}
    \end{bmatrix}, \quad
    E_2 = \begin{bmatrix}
0 & 0 & 0\\
0 & 0 & \sqrt{\gamma \delta t}\\
0 & 0 & 0
    \end{bmatrix}
\end{equation}
in the qutrit basis, for small time steps $\delta t$. This decay poses a fundamental limit for the fidelity of multi-qubit gate operations on neutral atom architecture \cite{saffmantrans}.\\

The second process we consider is \textit{erasure}, which encapsulates both atom loss and leakage where the qubit state leaves the computational basis through for example black-body radiation. It has been estimated that most errors in neutral atom quantum computers are a result of leakage compared to Pauli errors within the computational basis \cite{98erasure}. In coding theory, this is known as erasure, where the erasure symbol $ " ? " $ is now appended to the binary alphabet $\mathbb{F}_2 = (0, 1)$. It is known that (quantum) error correction codes benefit from erasure conversion, since erasure errors are easier to decode than $X$- or $Z$-type errors \cite{infotheory}. The decay width of leakage is denoted as $\omega$, and is an implicit function of the ambient temperature. We expect a tradeoff between radiative decay and leakage: the former imposing stricter limits on error rate thresholds than the latter. If erasure is detected using ancilla qubits, as proposed in Ref. \cite{iriscong}, we can define the \textit{erasure channel}
\begin{equation}
    \mathcal{L}^\text{erasure}(\rho) = (1-r)\rho\otimes\ket{0}\bra{0}+r\frac{\mathbb{I}}{2}\otimes\ket{1}\bra{1},
\end{equation}
dependent on an erasure probability $r$. Thus, we implicitly assume throughout this Article that we can detect and localise erasure in real time.

\subsection{Generating entanglement}

\begin{figure}[t]
    \centering{
    \includegraphics[width=0.4\textwidth]{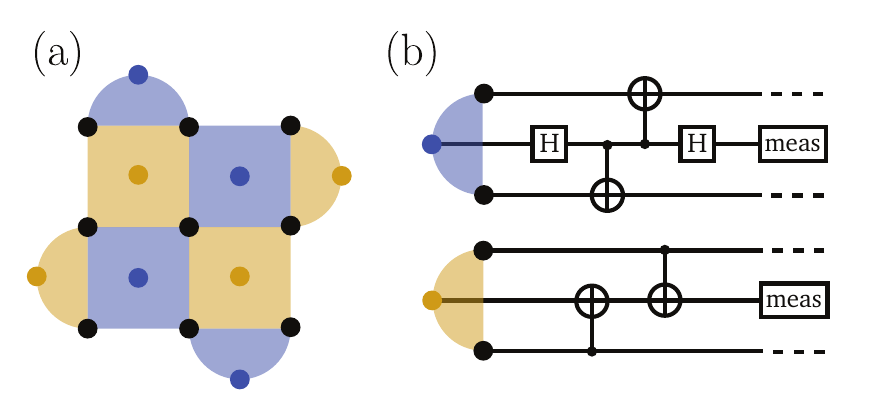}
    \captionsetup{justification=Justified}
    \caption{\textbf{(a)} The $[[9, 1, 3]]_2$ rotated surface code, where \textbf{black} circles denote data qubits that carry the 2 logical degrees of freedom, \textbf{\textcolor{xblue}{black}} faces denote $X$-stabilisers and \textbf{\textcolor{xgolden}{golden}} faces denote $Z$-stabilisers. \textbf{(b)} For each stabiliser type, the relevant circuit is shown for 2-plaquettes. $H$ is the Hadamard gate and \textit{meas} denotes a mid-circuit measurement. 4-plaquettes have analogous circuit designs.}
    \label{fig:ouwe}
    }
\end{figure}

We briefly review quantum error correction and the role of stabilisation. Topological codes are a class of codes whose topological properties dictate the error correcting capabilities of the code. Promising examples are the surface code and the toric code. Such codes are composed of two qubit types: \textit{data qubits} that carry the logical quantum information in non-local degrees of freedom, and \textit{stabilisers} that have to be read out to detect and recover errors. There are 2 sets of stabilisers: $\mathbb{S}_X$ detects phase errors and $\mathbb{S}_Z$ detects amplitude errors. CSS (Calderbank-Shor-Steane) codes are a family of quantum stabiliser codes \cite{CS, S}, generated by two classical linear codes $C_1$ and $C_2$ such that $C_2\subseteq C_1 \subseteq \mathbb{F}_q^n$ on the finite field of $q$ elements \footnote{$\mathbb{F}_q$ is the finite field of $q$ elements, also called the Galois field, where $q=p^s$ is an integer power of some prime number $p$. $\mathbb{F}_q^n$ is equivalent to the $n$-dimensional field $\mathbb{F}_q\times\cdots\times\mathbb{F}_q$.}. This CSS construction leads to a $q$-ary $[[n,k,d]]_q$-code, with length $n$, dimension $k$ and minimum distance $d$. Each underlying classical code generates a set of stabilisers that can correct for either one of the error types. A favourable property of surface codes is that interactions are \textit{local}, i.e. the qubits can be laid out in a planar graph and only require nearest-neighbour interactions within a given plaquette to engineer a QEC code, as seen in Fig. \ref{fig:ouwe} \footnote{Note that in this Article, we use a depiction of the $d=3$ rotated surface code to represent \textit{any} topological CSS codes, because of its simplicity.}.\\

Let $\mathbb{S} = \mathbb{S}_X \cup \mathbb{S}_Z $ be the group of stabilisers. Stabilisation between data qubits and stabiliser qubits $S\in\mathbb{S}$ is achieved through local 2-qubit entanglement operations. Neutral atoms provide a multitude of implementations, such as the simple constant pulses proposed by Jaksch et al. \cite{jaksch} or a time-optimal pulse proposed by Jandura et al. \cite{sven}, which will be central to this Article. \textcolor{black}{These are just a handful of examples of a very rich number of pulse protocols developed in the past few years, such as the pulse proposed by Pagano et al. \cite{alice} that achieves a pulse time very comparable to Ref. \cite{sven} and has a lower integrated Rydberg time compared to the protocol introduced in Ref. \cite{levine}. Recently, also a pulse that is robust against time-dependent variations in the laser control parameters was proposed by Mohan et al. \cite{madhav}.}\\

\begin{figure*}
    \centering{
    \includegraphics[width=0.8\textwidth]{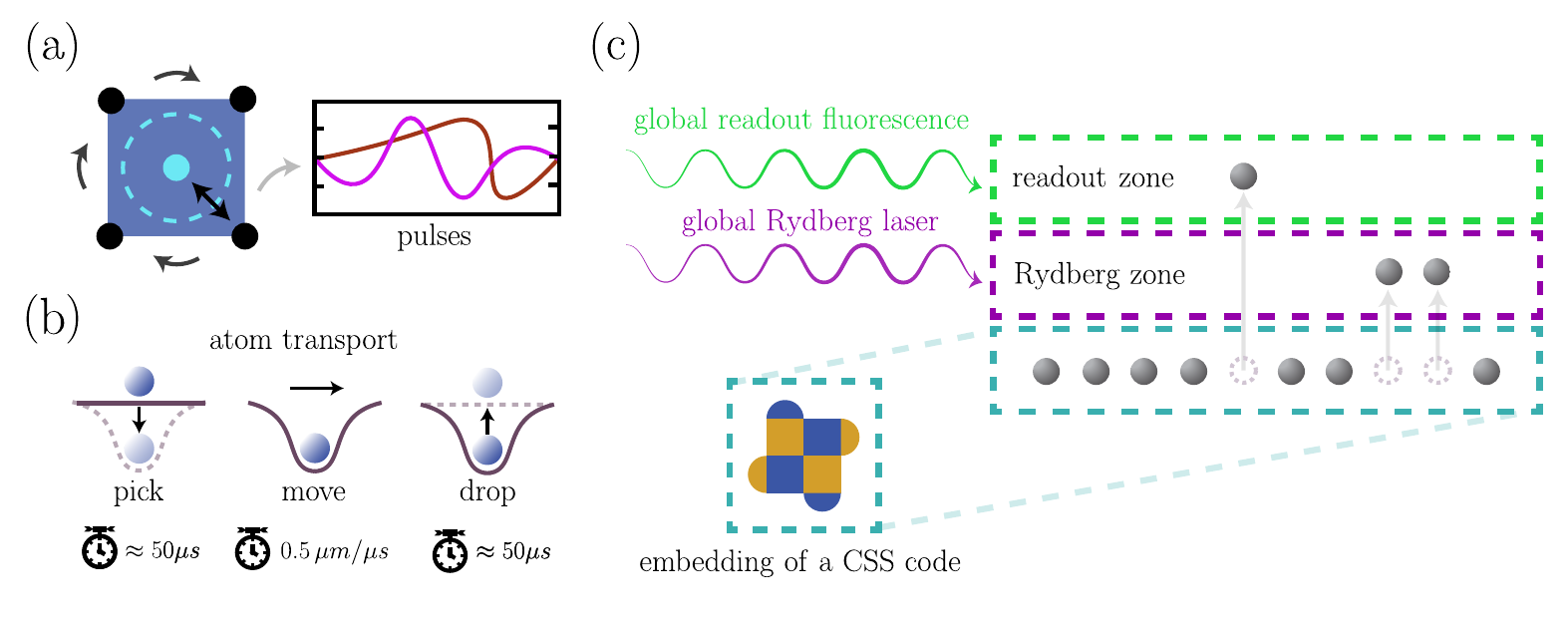}
    \captionsetup{justification=Justified}
 \caption{\textbf{(a)} Entanglement protocol for 4-plaquettes. The dashed circle indicates the Rydberg blockade radius, limited below the interatomic distance. Controlled logic between stabiliser and data qubits is performed by transporting the neighbouring qubits within the blockade radius, and applying an appropriate series of pulses before moving it back to its original position in the lattice. We subsequently entangle other data qubits in a clockwise fashion, as indicated by the grey arrows. \textbf{(b)} Visualisation of atom transport, with all relevant timescales indicated below \cite{movabletweezers, 60microns, meas}. AOM (acousto-optical modulator) tweezers have the ability to pick up an atom from its stationary trap and move it to a different location. Such transport allows for flexibility during execution of an algorithm, though transpilation may pose a serious challenge \cite{transpilationishard}. \textbf{(c)} Schematic architecture for neutral atom quantum computing experiments, tailored here for quantum error correction. We embed the CSS code in a 2D plane of atoms (grey zone), here seen from the side. Single-qubit operations are performed in the grey zone, entanglement is mediated in the purple Rydberg zone, and measurements of the qubit states are performed in the green readout zone. AOMs transport atoms between layers. The Rydberg laser that mediates the transition $\ket{1}\leftrightarrow\ket{r}$ is global, as well as the laser that reads out qubit states through fluorescence. Hence, dedicated zones are required to avoid crosstalk. Single-qubit operations in the grey zone are local and can be tuned from lattice site to site, within realistic experimental bounds.}
    \label{fig:atum}
    }
\end{figure*}

Fig. \ref{fig:atum}(a) provides an implementation scheme of 5-qubit entanglement on the surface code. We implement subsequent 2-qubit gates in a clockwise fashion, and assume no interaction between Rydberg-excited data qubits. Movable tweezers, depicted in Fig. \ref{fig:atum}(b), can transport atoms within their stabilisers' Rydberg blockade regime, where controlled logic can be performed, before moving them back to their initial positions in the lattice. Fig. \ref{fig:atum}(c) shows a complete schematic overview of how quantum error correction can be achieved on a neutral atom platform, showing an architecture with different dedicated zones for data storage, entanglement and readout.\\

Given is a time-dependent set of $2L$ laser controls $[z(t)]\in\mathbb{C}^L\times\mathbb{R}^L$ that implements a multi-qubit entanglement operation for time $t\in[0, T]$ on a set of $L$ qubits arranged on a lattice $\Lambda$. Here we focus on the subset of controls
\begin{equation}
    \{\Omega_i, \varphi_i, \Delta_i\}_{i\in\Lambda}\in\big([0,\bar{\Omega}]\times[0,2\pi)\times[-\bar{\Delta},\bar{\Delta}]\big)^{L},
\end{equation}
where $i$ denotes a qubit site, $\Omega e^{i\varphi}$ are Rabi frequencies of the transition between $\ket{1}$ and $\ket{r}$, and $\Delta$ are laser detunings of the $\ket{r}$-state, which are implicitly time-dependent functions. $\bar{\Omega}$ and $\bar{\Delta}$ are the maximum values for the control parameters that are experimentally feasible. Let the control Hamiltonian $\mathcal{H}[z(t)]$ be generated by the controls through the form
\begin{equation}
    \mathcal{H}[z(t)] = \sum_i \frac{\Omega_i(t)}{2}X_i + \sum_i \Delta_i(t) n_i + \mathcal{H}_{\text{drift}}. 
\end{equation}
where $X_i = \ket{1_i}\bra{r_i}+ \text{h.c.}$ addresses the single-photon transition between $\ket{1_i}\leftrightarrow\ket{r_i}$, $n_i = \ket{r_i}\bra{r_i}$ is the occupancy of the Rydberg state, and the drift Hamiltonian is given by the Rydberg-Rydberg interactions
\begin{equation}
    \mathcal{H}_\text{drift} = \sum_{i>j} \frac{C_6}{R_{ij}^6}\ket{r_ir_j}\bra{r_ir_j},
\end{equation}
summed over all Rydberg states $\ket{r_i}$ with interatomic distance $R_{ij}$ and van der Waals coefficient $C_6$. \\


We take a Trotterisation of the total unitary evolution in Fig. \ref{fig:ouwe}(b), chopping controls $[z(t)]$ into $N$ equidistant time slices of duration $\delta t = T / N$, and apply Kraus operators given in Eq. (\ref{eq:kraus}) at every time step. In the limit of $N\to\infty$, this would recover the exact Lindbladian evolution, though we can truncate for small finite $N$ and achieve sufficient accuracy. To find the diagonal $[\chi_{\mu\mu}]$ components, we consider the $\Gamma$-matrix defined by
\begin{equation}
    \Gamma = \sum_\mu \mathfrak{D}_\mu \otimes \mathfrak{D}_\mu^\dagger,
\end{equation}
that contains all relevant information about the propagation of errors throughout a quantum system through projections. If we  expand our Kraus operators in the 5-qubit Pauli basis
\begin{equation}
    \mathfrak{D}_\mu = \frac{1}{2^N} \sum_\nu \text{Tr}(\mathfrak{D}_\mu \mathcal{P}_\nu) \mathcal{P}_\nu,
\end{equation}
to obtain the $\mathcal{E}^{\text{twirl}}(\rho)$ channel (\ref{eq:twirl}), the diagonal $[\chi_{\mu\nu}]$ components follow consequently from the relation
\begin{equation}
    \chi_{\mu\nu} = \frac{1}{4^N} \text{Tr}\bigg[\bar{\Gamma} \cdot (\mathcal{P}_\mu\otimes \mathcal{P}_\nu)\bigg],
\end{equation}
over the modified $\Gamma$-matrix:
\begin{equation}
    \bar{\Gamma} = (\mathbb{I}\otimes U) \Gamma (U^\dagger \otimes \mathbb{I}),
\end{equation}
where we insert $U$ to compare to the noiseless evolution $\mathcal{F}(\rho)$. Because the readout time $\tau_\text{meas}$ is of much greater order than all relevant decoherence timescales, we can assume all Rydberg states have decayed before the full measurement is concluded \cite{meas}. Later on, we will see that this gravely affects the logical error rate bounds. More details of the precise step-by-step calculation are elaborated on in Appendix \ref{sec:chi}.

\section{Statistical model of errors}\label{sec:statmech}

The \textit{statistical-mechanical mapping} is a mathematical duality between the performance of quantum error correction codes and a specific family of random bond Ising models or random plaquette gauge models. This duality was first realised by Dennis et al. \cite{topologicalquantummemory} and later generalised and applied for correlated noise \cite{chubb, higgs, ookstat}. Because of the statistical nature of errors, it is very natural to borrow tools from statistical physics to analyse the behaviour of error propagation within certain noise models, and their impact on the overall performance of the code. This gives us a concrete way of predicting upper bounds on error rate thresholds without using any decoder, such as the Blossom algorithm \cite{blossom} or belief propagation \cite{belief}.\\

First, we address the gauge symmetry of quantum error correction briefly. An error configuration $e$ is said to be \textit{homologically trivial} if there exists a subset of stabilisers $S^\star\subseteq\mathbb{S}$ such that $e = \prod_{s\in S^\star}s$. An error configuration leads to logical failure if and only if the recovery $r$ leads to an operator $e\oplus r$ that does not commute with every single of the $2k$ logical operators of an $[[n,k,d]]_q$-code. In other words, a decoder fails if and only if it fails to identify the right homology class of the underlying error $e$. Let $\bar{e}$ be the equivalence class of errors that are in the same homology class as some error configuration $e$, i.e.
\begin{equation}\label{eq:homology}
    \bar{e} = \{e + S \; | \; S\in\mathbb{S} \}
\end{equation}
is the quotient group of $e$ over the stabiliser group $\mathbb{S}$. Trivial errors are in the same equivalence class as the 0-error. The statistical mapping must abide by this same gauge symmetry.\\

\begin{figure}[t]
    \centering{
    \includegraphics[width=0.5\textwidth]{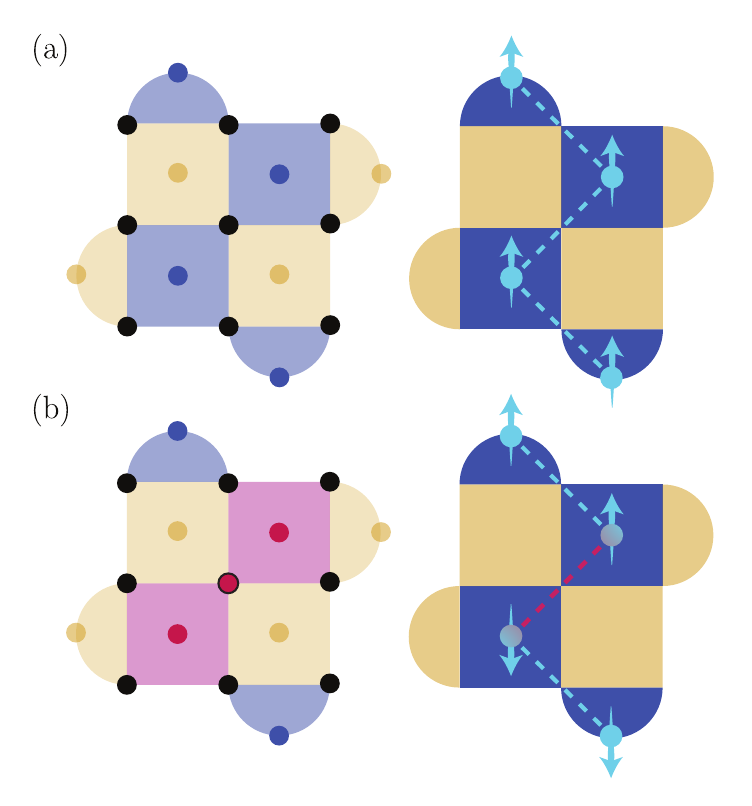}
    \captionsetup{justification=Justified}
    \caption{\textbf{(a)} A CSS code is mapped to a spin gauge model with random bonds, with spins $s^X$ (\textbf{\textcolor{xblue}{black}}) and $s^Z$ (\textbf{\textcolor{xgolden}{golden}}) representing stabilisers and bonds representing data qubits. Bonds between spins $s^i$ are ferromagnetic ($+J$) and favour long-range order, if the corresponding data qubit does not have an error that can be picked up by the stabiliser type $i$. Such bonds are depicted as blue dashed lines. \textbf{(b)} If a data qubit undergoes an error (red), triggering two flags (pink), its respective bond turns locally anti-ferromagnetic $(-J)$, indicated by a red dashed line.}
    \label{fig:spin}
    }
\end{figure}

Our focus is to find a classical $\mathbb{Z}_2$ gauge theory model whose thermodynamical properties mimic the behaviour of errors on topological codes. We assign a classical spin degree of freedom $s_i \in \mathbb{Z}_2 = \{\uparrow, \downarrow\}$ to each stabiliser qubit site, and map data qubits to bonds or plaquettes whose strength $(\pm J)$ is dictated by the underlying error configuration, as depicted in Fig. \ref{fig:spin}. The gauge spin model must satisfy a few properties:
\begin{enumerate}
    \item \textit{Topology compatibility} - It must be consistent with the underlying topology and boundary conditions of the quantum code. For instance, the dual Hamiltonian of the toric code must match its topology: $\mathcal{T}^2 = \mathcal{S}^1 \times \mathcal{S}^1$ for the 2D toric code and $\mathcal{T}^3 = \mathcal{S}^1 \times \mathcal{S}^1 \times \mathcal{S}^1$ for the 3D toric code.
    \item \textit{Configuration averaging} - The distribution of the random interactions must mimic error correlations as predicted by the twirled Pauli channel. The error probability $p$ determines the probability of a local anti-ferromagnetic coupling.
    \item \textit{Gauge symmetry} -  Errors which are equivalent to each other within the homology class Eq. (\ref{eq:homology}) must map to the exact same family of Hamiltonians. 
\end{enumerate}
We aim to find a dual statistical-mechanical Hamiltonian $\mathcal{H}(\vec{s}_i)$. Let the bond strength be defined as $\eta J$, with magnitude $J$ and probabilistic sign $\eta$ \footnote{The signum function $\eta$ is a function that tells us the sign of a quantity, such as a +1 coupling strength for no error, 0 for an erasure, and -1 for a Pauli error.}. Let $\pi(\bar{e})$ denote the probability that a certain error configuration belongs to the homology class of $e$. Then the \textit{Nishimori conditions} \cite{nishimori} are the set of constraints that yield the right Hamiltonian parameterisation such that 
\begin{equation}
    \mathcal{Z}_e[\eta, J] = \pi(\bar{e}),
\end{equation}
where $\mathcal{Z}_e$ is the partition function of the gauge Hamiltonian. The statistics match exactly since the Hamiltonian is invariant under the gauge symmetry generated by the stabilisers $S\in\mathbb{S}$.\\

At low temperature $T$, the system favours global order. Increasing $T$ allows more thermal fluctuations to disorder the system, up until a critical temperature $T_c$ above which all global order is broken. This 2$^{\text{nd}}$ order phase transition marks a phase boundary between order and disorder. The intersection point with the Nishimori condition yields the threshold error rate $p_\text{th}$ below which quantum error correction displays order, i.e. can drive the logical error rate arbitrarily close to 0 if the code distance $d$ is scaled up sufficiently high. The \textit{Nishimori point} where the Nishimori line crosses the phase boundary is a \textit{renormalisation group fixed point} \cite{fixedpoint}.\\

From now on, we make a distinction between 3 error types. \textit{Pauli errors} occur at a rate $p$, and signify errors within the computational basis. Three distinct types are considered: $X$ (amplitude) and $Z$ (phase) errors are chosen to be elementary errors; a $Y$ error is simply the occurrence of both types of errors at a single site: $Y = ZX$ \footnote{Note that we dropped the global phase $i$ here, which yields $Y^\dagger = -Y$.}. $M$ errors are \textit{measurement errors}, appearing at a rate $q$. We make this explicit distinction between spacelike errors and timelike errors to paint contrast between a (2+1)D and 3D quantum error correction code. Finally, we have \textit{erasure errors} denoted by the erasure symbol $" ? "$, at a rate $r$.

\subsection{$\mathbb{Z}_2$ Gauge theory - Random bond Ising model}

Here, we construct a dual Hamiltonian on a $d \times d$ lattice of data qubits, where $d$ is also implicitly the measure of the code distance of the topological code in which we encode our quantum information. Let $\mathcal{P}^1$ denote the set of single-qubit Pauli operators
\begin{equation}
    \mathcal{P}^1 = \{\mathbb{I}, X, Y, Z\},
\end{equation}
where $\mathbb{I}$ denotes the single-qubit identity operator (unless specified otherwise by a subscript index). Then, $\mathcal{P}^N$ denotes the $N$-qubit set of Pauli strings: $\mathcal{P}^N = \bigotimes_{n = 1}^N \mathcal{P}^1$. Let the scalar commutator $[[\cdot, \cdot]]: \mathcal{P}^{N}\times\mathcal{P}^{N}\to\mathbb{C}$ over two operators $O_1, O_2 \in 
\mathbb{C}^{2^N\times 2^N}$ be defined as
\begin{equation}
    [[O_1, O_2]] = \frac{1}{2^N} \text{Tr}\left[O_1O_2O_1^\dagger O_2^\dagger\right].
\end{equation}
Then, in its most general form, the dual Hamiltonian reads 
\begin{equation}\label{eq:chubb}
    \mathcal{H}_e[\vec{s}] = -\sum_{j, \sigma \in \mathcal{P}_i} J_j(\sigma) [[\sigma, e]] \prod_k [[\sigma, S_k]]^{s_k},
\end{equation}
where, $J$ determines the coupling strength, $[[\sigma, e]]$ is the signum fucntion that determines the sign of the coupling, and the product $\prod_k \cdots$ is the product of spins connected by the coupling \cite{chubb}. This Ising model has the gauge symmetry of the stabilisers built in, which is invariance under multiplication with elements of the stabiliser group:
\begin{equation}
    e \mapsto eS,
\end{equation}
for $S\in\mathbb{S}$, and generalises for any underlying error model. For CSS codes with independent $X$ and $Z$ errors, we can separate this Hamiltonian into two independent ones, pertaining to each error type. This \textit{gravely} simplifies our analysis.\\

We can generalise the Hamiltonian formalism for arbitrary dimensions. The $D$-dimensional isotropic partition function $\mathcal{Z}[J, \eta]$ dropping the label $e$, corresponding to a toric code with one single error type occurring at uniform rate $p$, is given by 
\begin{equation}\label{eq:z}
    \mathcal{Z}[J, \eta] = \sum_{\{\vec{s}\}} \exp{\left(\beta J\sum_{\langle i,j\rangle}\eta_{ij}s_i s_j\right)},
\end{equation}
where $\{\vec{s}\}=\{-1,1\}^{\times d^{D}}$ is the set of all possible spin configurations, $\sum_{\langle i, j \rangle}$ is the sum over nearest neighbour sites only, $\beta = T^{-1}$ is the inverse temperature, $\eta_{ij}\in\{-1, 1\}$ is the coupling between neighbouring spins $i$ and $j$, and $J$ is the interaction strength set to unity. It is easy to see that the $\mathbb{Z}_2$ gauge symmetry is given by
\begin{equation}
    s_i \mapsto \sigma_i s_i, \quad \eta_{ij} \mapsto \sigma_i \sigma_j\eta_{ij},
\end{equation}
for some set of gauge variables $\sigma_i\in\mathbb{Z}_2.$ 
In this model, we can define an order parameter $m^2$ that distinguishes the ferromagnetic and paramagnetic phases, given by the \textit{mean-squared magnetisation}
\begin{equation}\label{eq:mag}
    m^2 = \lim_{|i-j|\to\infty}\langle {s_i} {s_j} \rangle
\end{equation}
For $T<T_c$, $m^2 > 0$, and for $T>T_c$, $m^2 = 0$. The critical temperature $T_c$ that marks the phase boundary can be derived from a multitude of quantities, such as a divergence in the magnetic susceptibility.\\

For an independent uniform error model, the \textit{Nishimori condition} is given by
\begin{equation}\label{eq:nishimori}
    J = \frac{1}{2\beta}\ln{\left(\frac{1-p}{p}\right)}
\end{equation}
for a depolarisation probability $p$. Along the Nishimori line, the free energy of the ferromagnetic and anti-ferromagnetic regions are identical, yielding enhanced symmetry. For $D = 2$ and $D = 3$, Monte Carlo simulations \cite{nishimori, 3drbim} predict a second order phase transition for errors at constant rate $p$, at threshold error rates of
\begin{equation}
    p_{\text{th}}^{2\text{D}} \approx 0.109\quad \text{and}\quad p_{\text{th}}^{3\text{D}} \approx 0.250,
\end{equation}
which are in agreement with classical simulations of the toric code, such as those provided by the {\fontfamily{cmtt}\selectfont
stim} Python package \cite{stim}. The surface code has a lower threshold because the boundary conditions make it less likely for the decoder to correctly identify the right equivalence class of certain error configurations, and faithfully correct for them. Toric code simulations therefore provide an upper bound on the performance of the surface code.\\

\subsection{$\mathbb{Z}_2$ Gauge theory - Random plaquette model}\label{sec:plaq}

Quantum error correction experiments usually involve measuring syndromes over a multitude of \textit{cycles}. Since measurements are not guaranteed to produce the desired outcomes, either due to uncertainty in the measurement procedure itself or due to an amplitude error occurring on a stabiliser qubit after stabilisation, a number of cycles on the order of $d$ are ran to make sure that an error history can be faithfully traced back. Random plaquette gauge models map directly to such (2+1)-dimensional spatiotemporal quantum error correction codes \cite{higgs, kogut}. The corresponding Hamiltonian is a generalisation of Eq. (\ref{eq:chubb}) to 2+1 dimensions, such that 
\begin{equation}
    \mathcal{H}_e[\vec{s}] = -J\sum_P \tau_P \prod_{P\ni i} s_i,
\end{equation}
where $P$ are plaquettes with signs $\{\pm 1\}$ and $P\ni i$ are edges that are adjacent to the plaquette $P.$ The gauge transformation is
\begin{figure}[b]
    \centering{
    \includegraphics[width=0.45\textwidth]{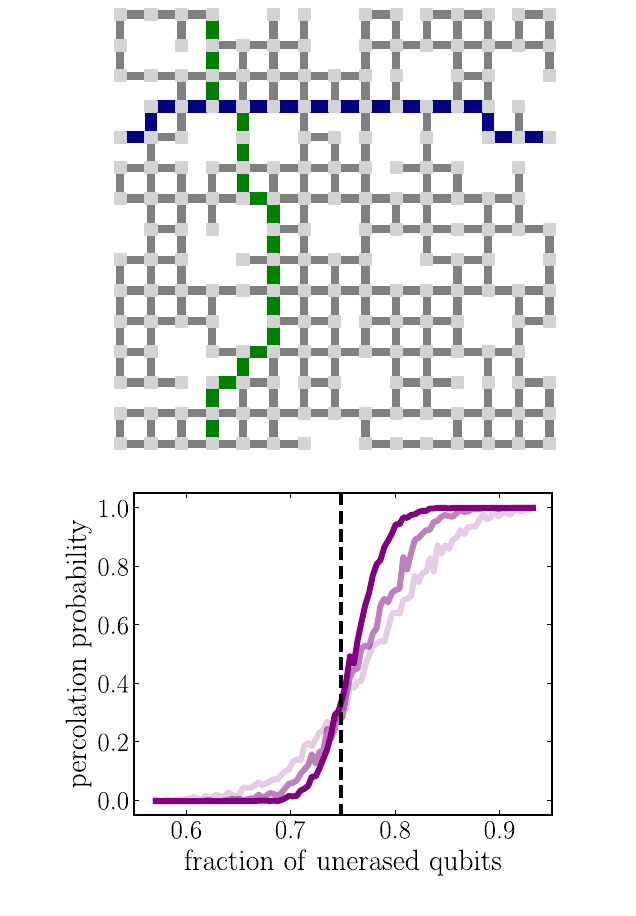}
    \captionsetup{justification=Justified}
    \caption{\textbf{(Top)} Example of percolation on a 15$\times$15 spin system that is dual to the toric code, where spins are given by grey nodes. Bonds and vertices are randomly removed at a uniform rate $r = 0.15$, leaving a subgraph on which top-bottom and left-right percolation can be found. Examples of logical support are indicated in \textbf{\textcolor{olive}{green}} and \textbf{\textcolor{cerulean}{black}} respectively. As periodic boundary conditions apply, percolation clusters must connect left-right and top-bottom on the same horizontal respectively vertical line. \textbf{(Bottom)} Percolation probability as a function of the fraction of unerased qubits, for system sizes $L = 5, 10, 25$. Darker colours indicate larger lattices. The vertical black line indicates the threshold at around 0.75 ($r \approx 0.25$.}
    \label{fig:perc}
    }
\end{figure}
\begin{equation}
    s_i \mapsto \sigma_i s_i, \quad \tau_{P} \mapsto \sigma_i \sigma_j \sigma_k \sigma_l \tau_P,
\end{equation}
for some set of gauge variables $\sigma_i\in\mathbb{Z}_2,$ and $\{i,j,k,l\}\in P$. The order parameter is given by the Wilson loop 
\begin{equation}
    \mathcal{W}[\tau] = \prod_{l\in\tau} s_l
\end{equation}
over a closed loop $\tau$ over the lattice, where $s_l \in \mathbb{Z}_2$ sit on edges. This quantity is gauge-invariant and non-local, providing an excellent candidate to quantify phase transitions, according to Elitzur's theorem \cite{elitzur}. Let $|\tau|$ be the length of the loop, and $S_\tau$ the enclosed area. In the ordered phase, the thermodynamical average of the Wilson loop is given by
\begin{equation}
    \langle\mathcal{W}[\tau]\rangle \propto \text{exp}\left(-c_1|\tau|\right),
\end{equation}
while in the disordered phase, we find 
\begin{equation}
    \langle\mathcal{W}[\tau]\rangle \propto \text{exp}\left(-c_2 S_\tau \right),
\end{equation}
where $c_1$ and $c_2$ are smooth functions of the temperature \cite{gauge, thatonephasediagrampaper}. These scaling laws are called the \textit{perimeter law} and \textit{area law} respectively. We stress that a 3D random bond Ising model does not describe measurement errors, (because we are implicitly solving $\partial\star u = 0$ for error chain 1-forms $u$, where $\partial$ is a total differential and $\star$ is the Hodge star operator, indicating we must map errors on data qubits to plaquettes). However, if there is no $z$-coupling, i.e. no measurement errors, it was proven by Suzuki \cite{fukinuke} that this model reduced to individual copies of a 2D random bond Ising model. In particular, a threshold error rate of
\begin{equation}
p_{\text{th}}^{(2+1)\text{D}} \approx 0.033
\end{equation}
was established for a uniform isotropic error model \cite{higgs}, at the intersection with the same Nishimori line generated by Eq. (\ref{eq:nishimori}).

\begin{figure*}[t]
    \centering{
    \includegraphics[width=0.75\textwidth]{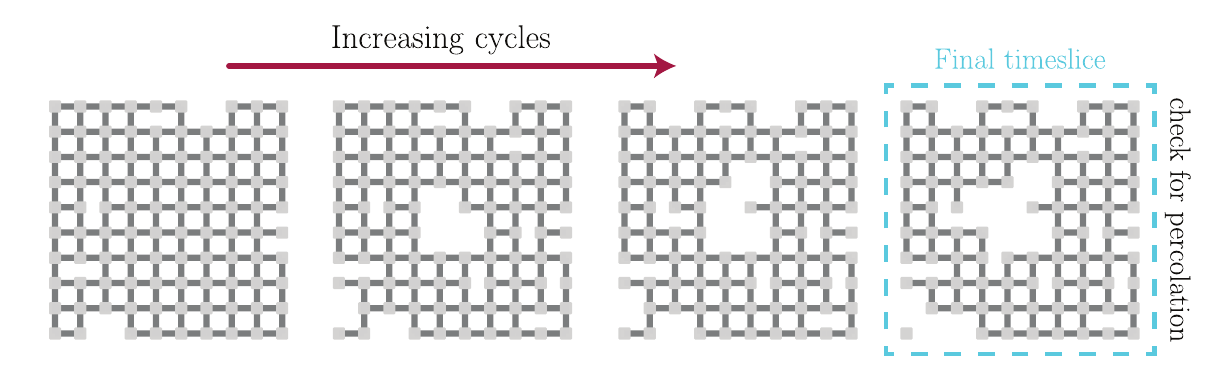}
    \captionsetup{justification=Justified}
    \caption{Example of the presence of erasure errors erasing data qubits (bonds) and stabiliser qubits (vertices) from the lattice. Time moves from left to right. Each panel shows the lattice at a later cycle at regular temporal intervals. Erased data qubits will remain erased throughout the entire duration of the quantum error correction experiment, though stabilisers may be refreshed depending on the availability of fresh atoms.}
    \label{fig:cycles}
    }
\end{figure*}

\subsection{Erasure model - percolation}\label{sec:perc}

Erasure of data qubits is equivalent to the problem of bond percolation, while erasure of stabiliser qubits is equivalent to site percolation. The erasure of stabilisers poses the most stringent danger to the existence of a logical qubit, since the erasure of a stabiliser automatically erases all of its bonds. Additionally, they are the only classical feedback we can infer from the system, so losing them gives us access to only part of the error history, affecting the threshold error rate.\\

Erasures do not corrupt quantum information if and only if all logical degrees of freedom are shielded from the effects of erasures, that is, neither bond nor site percolation is violated globally. In Fig. \ref{fig:perc}, an example configuration of percolation is displayed. Calculating the resulting error rate threshold $r_{\text{th}}$ is well known in literature and depends solely on the underlying geometry of the lattice \cite{perclist, 1/2percolation}. The tiling of a qubit lattice that embeds a topological quantum code will therefore affect this threshold. The order parameter associated to percolation is the percolation strength: the fraction of sites $\Pi$ that are a part of an infinite cluster:
\begin{equation}
    \Pi = \begin{cases}
      0 & \text{if}\; r < r_{\text{th}}, \\
      1 & \text{if}\; r > r_{\text{th}}.
    \end{cases} 
\end{equation}
Both left-right and top-down percolation are required for information to be protected on the toric code. In Fig. \ref{fig:cycles}, the temporal effects of erasure are depicted.

\subsection{Capturing correlations}
In order to insert the results from the Pauli twirling calculations, we can randomly draw local 5-qubit Pauli error configurations according to its underlying distribution function and apply them to all unit cells. Motivated by this structure, we partition the qubit lattice $\Lambda$ into a non-disjoint set of sub-lattices $\{\Lambda_i\}_i$ such that
\begin{equation}
    \Lambda = \bigcup_{i} \Lambda_i
\end{equation}
that are necessarily non-intersecting, i.e. $\Lambda_i \cap \Lambda_j \neq {\varnothing}$ does not imply $i = j$, which is a general requirement regardless of the code geometry and connectivity. If all sub-lattices were disjoint, then the model would resemble an independent error model after sufficient coarse graining. For every stabiliser $S_\mu\in\mathbb{S}$ we define the local neighbourhood as all data qubits that interact with the stabiliser, i.e. 
\begin{equation}
    \Lambda_{S_\mu} = \{\nu \; | \; H_{\mu\nu} = 1\},
\end{equation}
where $H_{\mu\nu}$ is the parity check matrix of the underlying CSS code. The Nishimori conditions for a general correlated error model are given by
\begin{equation}
    \beta J_i(\sigma) = \frac{1}{|\mathcal{P}_{\Lambda_i}|} \sum_{\tau\in\mathcal{P}_{\Lambda_i}} \ln{\phi_i(\tau)}[[\sigma,\tau^{-1}]],
\end{equation}
where the functions $J_i: \mathcal{P}_{\Lambda_i}\to\mathbb{R}$ are now defined for local neighbourhoods \cite{chubb}, which are the 5-qubit 4-plaquettes of the toric code, but can be generalised for any topological CSS code such as colour codes.\\

\begin{figure*}
    \centering{
    \includegraphics[width=0.99\textwidth]{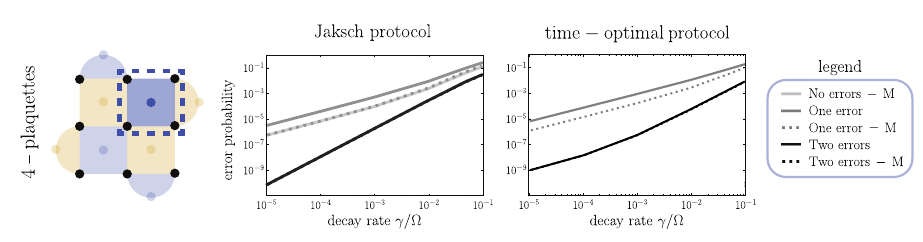}
    \captionsetup{justification=Justified}
    \caption{The error probability $p$ of a certain number of errors that can occur on 4-plaquette data qubits as a function of the decay width $\gamma$ in units of the Rydberg laser Rabi frequency $\Omega$, given by a sum of elements $\chi_{\mu\mu}$ from Eq. (\ref{eq:chi}). Here, no leakage or erasure is assumed. \textbf{M} in the legend indicates that a measurement error occurred (on the stabiliser qubit). Simulations were performed by using the finite element Kraus evolution for both pulse protocols, with $N=20$ Trotterisation steps for every single 2-qubit entanglement pulse. In the small error limit $\gamma/\Omega \ll 1$, we retrieve scaling laws compatible with heuristic arguments as highlighted in the main text. For 2-plaquettes, we retrieve similar plots and scaling laws, omitted for clarity's sake. For either protocol, the curves for '$2$ errors' and '$2$ errors + measurement error' are overlapping. For the time-optimal protocol, the curves for '1 error' and 'no errors + measurement error' overlap. The latter protocol is also completely symmetric with regards to $X \leftrightarrow Z$.}
    \label{fig:pulses}
    }
\end{figure*}

\section{Code performance analysis}\label{sec:codeperformance}

Now we analyse the performance of the toric code over a range of decay rates $\gamma$ and $\omega$, and quantify how the QEC error rate threshold depends on their interplay. For several discrete points in the $(\gamma, \omega)$-parameter space, we will calculate the error rate threshold $p_{\text{th}}$. The twirled channel $\mathcal{E}^{\text{twirl}}(\rho)$, as given in Eq. (\ref{eq:twirl}), gives us the likelihood of certain local error configurations as well as the likelihood of a measurement error with probability $q$. The random plaquette gauge model described in Sec. \ref{sec:plaq} allows us to evaluate the performance of QEC codes over different cycles $c$ by applying the twirled quantum channel
\begin{equation}
    \mathcal{E}^{(c)}(\rho) = \overbrace{\mathcal{E}^{\text{twirl}}\circ\cdots\circ\mathcal{E}^{\text{twirl}}}^{c\; \text{times}}(\rho).
\end{equation}
To effectively combat measurement errors, we perform $d$ rounds of measurements for a $d\times d$ toric code. Additionally, it is common practice to assume the final round of measurements is perfect, by reading out every data qubit and inferring a parity syndrome from that readout. Because neutral atom quantum computers typically suffer from long measurement timescales, the measurement time $\tau_{\text{meas}}$ dictates the clock speed of error correction. Since decoherence times are typically of lower orders than $\tau_{\text{meas}}$, we assume \textit{all} Rydberg states have emptied at the end of every cycle, which drastically impacts the influence of the CNOT entanglement protocol on the error rate threshold.\\

The rest of this Section is dedicated to obtaining qubit fidelities for the toric code implemented on a neutral atom quantum device, assuming two major loss channels mediated by the fragile Rydberg state. We evaluate the local error channels for 4-plaquettes, extracting probabilities $\chi_{\mu\mu}$ for certain twirled Pauli error strings on the data qubits, which includes the probability $q$ of a false flag error on the stabilisers, and insert these into a 3D random plaquette gauge model. Additionally, erasure rates are incorporated, whose effects are a result of the integrated Rydberg lifetime of the qubits plus the finite lifetime of the tweezer traps. Because of the anisotropy of our system, spacelike and timelike Wilson loops have different sensitivities to variations in temperature. Our calculations have shown that disorder is more spacelike than timelike, so that we sample timelike Wilson loops only. The details of the Metropolis algorithm and handling phase transitions are presented in Appendix \ref{sec:montecarlo} \cite{FSS, FSS2}.\\

\subsection{Twirled probabilities}

Under the Pauli twirling approximation, we can find probabilities associated to a certain Pauli error $\mathcal{P}_{\mu}$ occurring on a local cluster of 5 qubits, by calculating the diagonal terms $\chi_{\mu\mu}$, for different pulse implementations. The two implementations we consider in this Article are the Jaksch protocol \cite{jaksch} (also colloquially referred to as the $\pi$-$2\pi$-$\pi$ pulse) and the time-optimal Jandura protocol that minimises the Rydberg population time \cite{sven}. We inherently assume in our models that non-leakage decay from the Rydberg state goes to the $\ket{1}$-state, so that errors during the $CZ$-implementation can only corrupt phase information. Thus we find that $X$-stabilisation yields $X$-errors only, and likewise for $Z$-stabilisation. This way, we split our analysis up into a competition between independent $X/Z$ Pauli errors versus erasure errors.\\ 

Because of the gauge symmetry of topological quantum error correcting codes under the group of stabilisers $\mathbb{S}$, we simplify the analysis by grouping probabilities in terms of error count. For data qubit errors on general $n$-plaquettes, this corresponds to grouping
\begin{equation}
    \mathbb{P}(k \;\text{errors}) \;\; \text{and} \;\; \mathbb{P}(n-k\; \text{errors}),
\end{equation}
since they are congruent to each other under applying the stabiliser operator associated to the ancilla qubit that stabilises them.
The results are displayed for 4-plaquettes in Fig. \ref{fig:pulses}. A crucial feature of these simulations is that the measurement error rate is usually on the same order of magnitude as non-measurement error rates. This means that in the limit of very strong decoherence $\gamma\Omega \approx 1$, the underlying spin model starts to resemble more a 3D isotropic random plaquette gauge model than a 2D random bond Ising model, with a suppressed error rate threshold.\\

\begin{figure}[t]
    \centering{
    \includegraphics[width=0.45\textwidth]{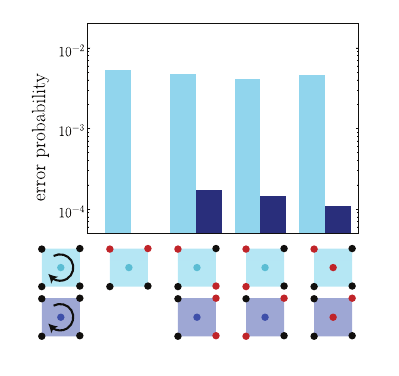}
    \captionsetup{justification=Justified}
    \caption{Elaborate example of different 2-qubit correlations on a 4-plaquette, for the time-optimal protocol at $\gamma/\Omega = 0.1$. The order of stabilisation starts from the top left and rotates clockwise. (Light)blue bars indicate the correlations between erroneous red qubits on the (light)blue plaquettes. Note that the first blue bar is missing because we only look at causal correlations between the second qubit being erroneous, and any qubit later in the stabilisation process being erroneous as well.}
    \label{fig:correl}
    }
\end{figure}

Importantly, from Fig. \ref{fig:pulses} we retrieve fidelity behaviour that is consistent with heuristic scaling laws. For the Jaksch protocol, the curves follow the 
\begin{equation}
    \lim_{\gamma\to 0} p = a\gamma^{F}
\end{equation}
scaling law at low $\gamma$-values for some $a\in\mathbb{R}^+$ and $F\in\{1,2\}$ denoting how many data errors occurred. This is consistent with the intuition that 1 decay process should occur with probability proportional to $\gamma$, while 2 decay events occur at a rate proportional to $\gamma^2$. For the time-optimal protocol, the tails of one- and two-error probabilities both scale with $\gamma$. The intuitive picture behind this scaling is that since we drive collective multi-qubit excitations to the Rydberg state, double errors occur at a rate $1 - (1-\gamma\delta t)^2\sim\mathcal{O}(\gamma)$ for small timescales $\delta t$.\\

\subsection{Handling erasure}

For the erasure rate $\omega$, we consider two loss processes. Leakage, as described in Sec. \ref{sec:atoms}, is a result of black body radiation (BBR) mediated transitions. Qubits can be erased with a probability proportional to the time spent in the Rydberg state and the black body scattering width. The standard value we adopt is $\Gamma_{\text{BBR}} = 2\pi\cdot 840$ Hz. Another loss process is a consequence of the finite lifetime of the trap, which is the dominant mechanism behind erasure. Traps are described by a lifetime $T_\text{trap}$, with an ensemble of atoms knowing exponential decay. Assuming a measurement time on the order of $\tau_\text{meas} \approx 1-10$ ms, and a trap lifetime on the order of $T_\text{trap} \approx 10-50$ s, we have a trap ejection rate that increases monotonically as a function of the number of cycles $c$.\\

One \textit{major challenge} of the statistical mapping is calculating Wilson loop averages on a 3D plaquette model with percolation. The loop $\tau$ cannot contain any erased qubits as that would automatically set its expectation value to 0, and the enclosed area $S_\tau$ is calculated by taking into account the geometry of the holes that are encircled by the loop. For higher erasure rates, the shape of Wilson loops becomes more convoluted, marking a major roadblock in our computations for high $r$. For this reason, we extrapolate to $\gamma\to 0$, knowing that for the final timeslice, percolation on a random bond Ising model in 2D gives thresholds at $r=0.5$ and $r\approx 0.25$ for bond and site-bond percolation respectively, the latter which we established in Fig. \ref{fig:perc}. At the final timeslice we must have logical support for both logical qubits. We can imagine two scenarios:
\begin{itemize}
    \item Lost qubits are irrevocably lost.
    \item Lost stabiliser qubits are refilled in the next cycle with atoms pulled from a fresh reservoir, which is maintained and monitored throughout the entirety of the experiment to ensure a near 100\% filling rate. Data qubits will never be refilled since erasures pose less stringent bounds on QEC than Pauli errors.
\end{itemize}
The latter will of course benefit QEC error rate thresholds, at the cost of more qubit resources and continuously monitoring a reservoir in parallel.


\begin{figure*}
    \centering{
    \includegraphics[width=\textwidth]{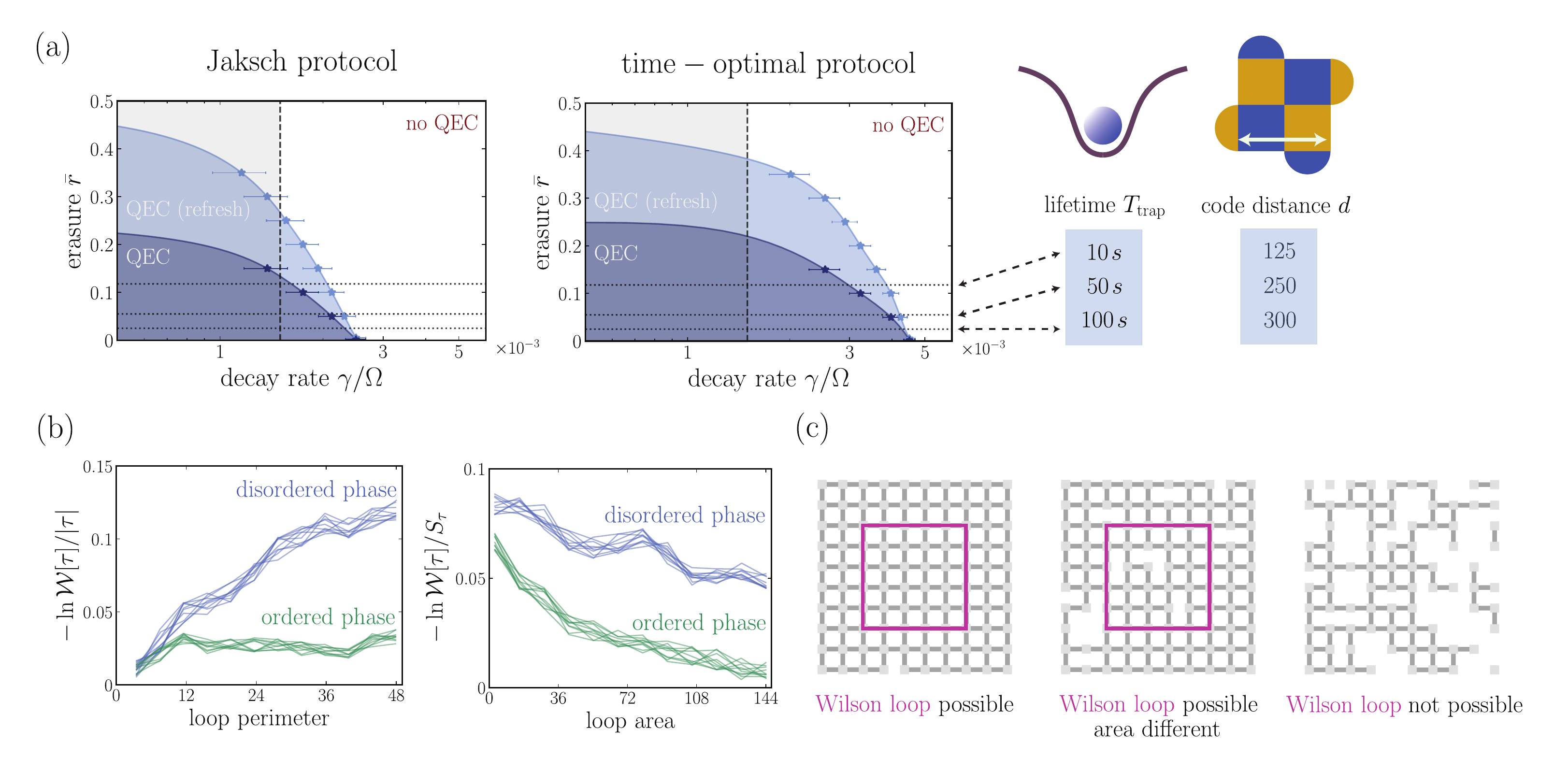}
    \captionsetup{justification=Justified}
    \caption{\textbf{(a)} The phase diagrams for our quantum error correction implementation of the toric code, with stabiliser circuits being implemented by either the Jaksch or time-optimal protocol. The dark blue regions indicate the QEC regime where arbitrarily low logical qubit fidelities can be achieved, granted $d$ is sufficiently large. The light blue regions indicate regions where QEC is only viable if lost stabilisers are replenished from some fresh reservoir, alleviating some of the fidelity bounds. The blank regime entails all parameters $(\gamma, \bar{r})$ such that QEC is no longer possible. \textcolor{black}{Error bars are a combined results of \textit{both} standard Monte Carlo finite sampling errors for 2500 error distributions and 10.000 steps, \textit{and} the uncertainty in $\gamma$ due to randomly distributing errors over a partially erased lattice yielding a different effective error rate standard deviation}. Because Wilson loop calculations for $r\gtrapprox 0.35$ are very difficult to perform in a Monte Carlo setting, we use a fit to approximate the phase boundaries, using the knowledge that for the limiting case of erasure-only errors ($\gamma\to 0$), we must recover 2D percolation thresholds calculated in Sec. \ref{sec:perc}. \textcolor{black}{A set of realistic system parameters are highlighted in black dashed lines/grey regions as examples.} \textbf{(b)} Typical graph of \textcolor{black}{the behaviour of} Wilson loop averages, normalised by perimeter/area. Each line is the average of 10 randomly chosen runs \textcolor{black}{averaged over 10.000 steps each}, at $r=0$ and $\gamma/\Omega = 10^{-3}$, for both the ordered ($T<T_c$) and disordered ($T>T_c$) regime, displaying a clear transition from a perimeter to an area law. The left graph shows $-\ln{\mathcal{W}[\tau]/|\tau|}$ \textcolor{black}{on the $y$-axis}, while the right shows $-\ln{\mathcal{W}[\tau]/S_\tau}$. \textbf{(c)} Examples of when Wilson loops can be easily constructed. If the loop encloses no erased qubits, the enclosed area $S_\tau$ is equal to a Euclidean area. If some erasures are enclosed, then the area is effectively less. For larger $\bar{r}$, the calculation of the area/perimeter law for arbitrary loops becomes convoluted and imprecise.}
    \label{fig:phase}
    }
\end{figure*}

\begin{table}[b]
\caption{Table of realistic experimental parameters. These values are not fixed, but serve as a practical reference throughout the remainder of this Article.}\label{tab:params}
\centering
\begin{tabular}{||c c c c||} 
 \hline
 Parameter & Value & Units & Description\\ [0.5ex] 
 \hline\hline
 $\Omega$ & $5-30$ & MHz & $-$\\
 \hline
 $\gamma$ & $96.5$ & $\mu$s & Lifetime of the $61s$ state \cite{madhav}.\\
 \hline
 $\Gamma_\text{BBR}$ & $5.3$ & kHz & All BBR transitions for $n\in [45,75]$ \cite{madjarov}.\\
 \hline
 $T_\text{trap}$ & 10-50 & s & $-$\\
 \hline
 $\tau_\text{meas}$ & 1-10 & ms & $-$\\
 \hline
\end{tabular}
\end{table}

\subsection{Logical qubit fidelity and lifetime}

First, we provide rough bounds on the lifetimes of logical qubits as predicted by our model. We adopt a set of parameters as summarised in Tab. \ref{tab:params}. The most dominant factor that impedes arbitrarily long quantum memory coherence times is erasure. Because lost data qubits cannot be replaced, the percolation threshold marks the absolute end of the logical qubit. Taking into account both atom loss and leakage, with a probability proportional to the integrated Rydberg time, we estimate the number of cycles that an experiment can maintain before losing the logical qubit. The total erasure rate is given by
\begin{equation}
    \omega = f_\text{int}\Gamma_\text{BBR}+\frac{1}{T_\text{trap}},
\end{equation}
where $f_\text{int}$ is the fraction of the integrated Rydberg time compared to the time it takes to conduct 1 cycle. We adopt this convention for later purposes. For infinitely stable traps ($T_\text{trap}\to\infty)$, the maximum number of cycles $c^\star$ is bounded by BBR-induced scattering during excitations to the Rydberg state, and will scale like $c^\star \sim 1/\Gamma_{\text{BBR}}.$ In this theoretical case, we expect a lifetime upper bound of around $\approx 3.5\cdot 10^3$ cycles. From trap lifetimes of the order of minutes, we already see that the lifetime of a logical qubit outlasts the physical qubit lifetimes. Therefore, we can establish that erasure plays only a minor role in the lifetime of logical qubits as long as traps are sufficiently long-lived.\\

Sweeping over $(\gamma, \bar{r})$, we retrieve a set of phase diagrams, displayed in Fig. \ref{fig:phase}, for the two pulse implementations considered. The diagrams represent the intersections of the \textit{Nishimori sheets} and the $p-T$ diagrams generated by sweeping over a range of $p$ and locating its associated critical temperature $T_c$. The Nishimori sheet is calculated using Eq. (\ref{eq:chubb}) and equates to
\begin{equation}
    T(p, r) = \frac{|\mathcal{P}|}{\sum_{\tau_{\text{Pauli}}\neq 0} [[\sigma, \tau^{-1}]]\text{ln}(\phi_j(\tau)(1-r))},
\end{equation}
under the identification that an erasure error is given by $\tau^{-1} = \textbf{0}.$ The effect of the erasure error rate $r$ drops out of the Nishimori condition through cancellation of logarithms, which makes sense physically considering it defines the distribution of wrong sign plaquettes on the plaquette gauge model, which is only well-defined on plaquettes that haven't yet been erased. \\

In the Figure, we have rescaled the $r$-axis and denote with $\bar{r}$ the \textit{effective erasure probability}, defined by the relation
\begin{equation}
    \bar{r} = 1 - \text{exp}\left(-\omega\tau_\text{meas}c\right),
\end{equation}
so that our phase diagrams are agnostic to measurement duration and the precise number of cycles/code distance $d$. The interpretation of this parameter is the average fraction of lost qubits at the \textit{final timeslice} of the experiment, which is what we must compare to $r_\text{th}$ calculated in Sec. \ref{sec:perc}. By adopting this parameter, we can change the erasure probability and the number of cycles without having to recalculate the diagram for these new specific parameters. \\

Fig. \ref{fig:phase} reveals plenty about the consequences of the underlying neutral atom physics to large-scale quantum error correction. One aspect that is clearly captured in these results is that erasure conversion is indeed a favourable protocol for QEC, as evident from the asymmetry of the diagrams. For more information, we refer the reader to Ref. \cite{conversion, conversion2}. For $^{88}$Sr, we have highlighted some experimental parameters for trap lifetimes in dashed lines, showing that strontium-based neutral atom quantum computers are already suitable for QEC experiments pertaining to quantum memory on toric codes. We find thresholds in the no-erasure limit of
\begin{equation}
    \gamma^\text{Jaksch}_\text{th} \approx (2.5 \pm 0.2) \cdot 10^{-3} \Omega \quad \text{and} \quad \gamma^\text{TO}_\text{th} \approx (4.5 \pm 0.2) \cdot 10^{-3} \Omega,
\end{equation}
which are consistent with literature on optimising neutral atom laser protocols for stabilisation for QEC \cite{sventwirl}, as it provides an upper bound on error rate thresholds without requiring an (optimal) decoder. Note that the results agree very well because of the simplicity of the toric code, having \textit{local} connectivity, and show compatibility with the range of state-of-the-art experimental parameters
\begin{equation}
    10^{-4}\Omega\lessapprox\gamma\lessapprox 10^{-3}\Omega.
\end{equation}
\textcolor{black}{The errors in $\gamma$ from our simulations are of a higher order of magnitude than those produced by simulations such as provided by the {\fontfamily{cmtt}\selectfont
stim} Python package, though we stress that the rough order or magnitude of an error rate threshold is more important than a $\sim$5\% variance. For more complex topological codes with more intricate error models, however, these minor Monte Carlo errors may be acceptable trade-offs, as our model may achieve greater time efficiency compared to performing many rounds of decoding which can be slow and underestimate the maximum error rate threshold.}\\

Using the phase diagrams, one can check for \textit{any} given set of experimental parameters whether QEC can be achieved, and what the lifetime of the qubit is with respect to the maximum number of measurement cycles $c$ (which is equal to the maximum code distance $d$). Of course, the model can be extended to study any distance-$d$ topological CSS codes\, and even to codes where the number of cycles is different from the code distance, to allow for a more precise characterisation. More generally, our model allows for \textit{any pulse protocol} to be studied under any noise type.

\section{Summary}\label{sec:summary}
Physical implementations of quantum error correction (QEC) on real hardware facilitate the propagation of correlated errors throughout the lattice of qubits in which the QEC code is embedded. This effectively lowers the threshold error rate $p_\text{th}$, below which increasing the code distance $d$ will suppress the logical error rate. In this Article, we have adopted the statistical mapping to characterise the entanglement strength of the propagation of errors for a neutral atom quantum computer, based on the level scheme of $^{88}\text{Sr}$, adopting physical implementations of laser pulses that realise multi-qubit gates. We unified this statistical model with the percolation model of erasures to obtain one overarching model that can handle all error types simultaneously. This resulted in a set a phase diagrams quantifying the interplay between erasure and Pauli errors, which can be used to gauge the effectiveness of QEC for a given set of experimental parameters.\\

We observed that Rydberg-related errors on a neutral atom system, under the assumption that the measurement time $\tau_{\text{meas}}$ is of larger magnitude than typical decoherence times $\tau_{\text{dec}}$, typically introduce more spacelike errors than timelike errors, so that the corresponding statistical model resembles more a 3D plaquette gauge model than a 2D depolarisation model with long-range spatial error correlations. We have also shown than time-optimal pulses give slightly more leeway for QEC than simple pulses do such as a $\pi-2\pi-\pi$ pulse, proving indeed they are more suited for QEC implementations, though the difference may only be sufficiently relevant for NISQ-era experiments.\\

In our model, similar to Ref. \cite{iriscong}, we assume that erasure can be tracked in real time through the use of ancilla qubits. This introduces a qubit overhead on the order of the number of physical qubits $n$. We can not exclude that in the future, more efficient overhead would be possible. Moreover, we have only studied the error rate thresholds in this Article; future work could also focus on the exponential suppression of the logical error rate.\\

Further research could investigate more extensive error models, including sources like stray fields or classical laser noise, the latter of which was studied in Ref. \cite{sse}. Furthermore, one can investigate the effect of simultaneous 5-qubit stabiliser pulses on code performance; for more discussion on this we refer the reader to Ref. \cite{sventwirl}. Lastly, our analysis can be generalised to systems where data qubits interact through their mutual Rydberg-Rydberg interactions, so that spatial correlations are more prone to occur.

\section*{Acknowledgements}
We thank Leo Radzihovsky, Jasper van de Kraats, Robert de Keijzer, Raul Parcelas Resina dos Santos, Denise Ahmed-Braun and Emre Akaturk for fruitful discussions. This research is financially supported by the Dutch Ministry of Economic Affairs and Climate Policy (EZK), as part of the Quantum Delta NL programme, and by the Netherlands Organisation for Scientific Research (NWO) under Grant No. 680.92.18.05. This project is partially funded by the European Union through the Horizon Europe programme HORIZON-CL4-2021-digital-emerging-01-03 via the project 10170144 (EuRyQa).

\section*{Data Availability}
The data that support the findings of this study are available from the corresponding author upon reasonable request. Simulations were performed using the Python packages {\fontfamily{cmtt}\selectfont
QuTiP} \cite{qutip}, {\fontfamily{cmtt}\selectfont pytorch}  Deep Learning library for GPU \cite{torch}, and {\fontfamily{cmtt}\selectfont pymatching} \cite{pymatching}.

\section*{Author declarations}
\subsection*{Conflict of interest} 
The authors have no conflicts to disclose.

\appendix

\section{Twirling}\label{sec:twirl}

Twirling is an approximation that takes an $n$-qubit error channel $\mathcal{L}: \mathbb{C}^{2^n\times 2^n}\to\mathbb{C}^{2^n\times 2^n}$ and keeps only those terms such that there exists a basis transformation that makes the channel diagonal in the transformed basis. More concretely, for a discrete set of unitaries $\mathcal{U}$, it is formally defined as the channel
\begin{equation}
    \text{tw}: \rho\mapsto \sum_{U\in\mathcal{U}}p_U U\rho U^\dagger.
\end{equation}
For the Pauli twirling approximation, this corresponds to a physical insertion of Pauli operators, drawn at random on random locations within the circuit. It is not known exactly how accurate the twirling approximation is, but it has been employed often with excellent agreement with exact simulations of the full system. It is known that Pauli twirling reduces the density matrix to the space of 1-designs. Another option would be to invoke Clifford twirling, which forms a 3-design. However, the physical insertions of unitaries from the Clifford group $\mathcal{C}^{(2)}$ would cause our error propagation model to break down. These twirls are useful in efficient simulations nonetheless since it has been shown that random circuits with local connectivity are approximate $t$-designs and can be simulated classically using only polynomial time.

\section{Calculating the entries of $[\chi_{\mu\nu}]$}\label{sec:chi}

We calculate the entries of $[\chi_{\mu\nu}]$ by constructing the $\Gamma$-matrix defined on the product of $n$-qudit spaces, and tracking at every time step $[t, t+\delta t]$ how the $\Gamma$-matrix transforms under the action of the unitary evolution $\mathcal{F}(\rho)$ and the complete set of Kraus decay operators $\{\mathfrak{D}_\mu\}_\mu$. Afterwards, using the completeness relation of the trace inner product 
\begin{equation}
    \langle\;\cdot\;, \mathcal{P}_\mu\rangle = \frac{1}{|\mathcal{P}_\mu|}\text{Tr}[\;\cdot\; \mathcal{P}_\mu],
\end{equation}
we can recover $[\chi_{\mu\mu}]$, the probability for a given Pauli error $\mathcal{P}_{\mu}$ within the Pauli twirling approximation.
In order to track this $\Gamma$-matrix, we initiate a $\Gamma_0$-matrix in the identity state of both subspaces $\Gamma_0 = \mathbb{I}_n^{\otimes 2}$, and for subsequent timeslices we update it according to the scheme
\begin{equation}
    \Gamma_{t+1} = \sum_\mu \mathfrak{D}_\mu (U_t \otimes \mathbb{I}) \Gamma_t (\mathbb{I} \otimes U_t^\dagger) \mathfrak{D}_\mu^\dagger,
\end{equation}
which, for $t\to N$, retrieves the final state $\Gamma_N = \Gamma$. We do not care about $\Gamma$ itself, however, but rather how it captures the physics of decoherence, relative to the unitary noise-free evolution. To find $\Gamma$ relative to the coherent evolution channel $\mathcal{F}(\rho)$, we transform to the Heisenberg picture under the unitary action of $U$, and find 
\begin{equation}
    \Gamma = \sum_\mu \mathfrak{D}_\mu (U_n \otimes U) \Gamma_{N-1} (U^\dagger \otimes U_n^\dagger) \mathfrak{D}_\mu^\dagger.
\end{equation}
On neutral atom quantum computers, measurement time is still the dominant timescale of a quantum error correction experiment, with $\tau_{\text{meas}}\approx 10$ ms being the order of magnitude for non-destructive fluorescence readout. Therefore, we can assume that during measurement, all remaining Rydberg states will have decayed back to the $\ket{1}$-states (or erased). Choosing $\bar{\delta t} \gg 1 / \gamma$, and $\bar{\mathfrak{D}}_\mu = \mathfrak{D}_\mu( \bar{\delta t})$, the final $\Gamma$-matrix becomes
\begin{equation}
    \Gamma = \bar{\mathcal{D}}_\kappa\left(\sum_\mu \mathfrak{D}_\mu (U_n \otimes U) \Gamma_{N-1} (U^\dagger \otimes U_n^\dagger) \mathfrak{D}_\mu^\dagger\right)\bar{\mathcal{D}}_\kappa^\dagger.
\end{equation}

\section{Monte Carlo details}\label{sec:montecarlo}

\subsection*{Metropolis algorithm}

Finding the critical temperature $T_c$ that signifies the 2$^\text{nd}$ order phase transition can be done in a multitude of ways. Here, we trace thermodynamical quantities $\langle\mathcal{X}\left[\vec{s}\right]\rangle$, where brackets denote thermal averages over error configurations, which are defined by the Boltzmann sum
\begin{equation}\label{eq:therm}
    \langle \mathcal{X}[\vec{s}]\rangle = \sum_e \pi(e) \sum_{\vec{s}} \frac{\exp{\left(-\beta\mathcal{H}_e[\vec{s}]\right)}}{\mathcal{Z}_e} \mathcal{X}[\vec{s}].
\end{equation}
for arbitrary functions $\mathcal{X}[\vec{s}]$. The sum over spin configurations $\sum_{\vec{s}}$ is handled by the Metropolis algorithm, while the sum over error configurations $\sum_e \pi(e)$ is mitigated by truncating error configurations with a probability smaller than a threshold value. The inverse temperature plays the role of the probability that thermal fluctuations dominate the system. In order to evaluate thermal ensemble averages of the form Eq. (\ref{eq:therm}), we use the Metropolis algorithm. A random spin configuration $\{\vec{s}^{\;0}\}$ with initial energy $E_0$ is initialised. Then, spins are randomly flipped to obtain a new trial state, whose energy $E_t$ is compared to the previous iteration's. We accept the thermal fluctuation if
\begin{equation}
    \Delta E < 0 \quad \text{or} \quad \exp{\left(-\beta\Delta E\right)}\geq r
\end{equation} 
for a random scalar $r\in(0,1)$. We repeat until sufficient convergence is achieved. Let $\text{supp}(e)$ be the support in $e$, i.e. the total number of single-qubit errors that are contained in the configuration $e$. If errors occur at a uniform rate $p$, such as in a depolarisation model Eq. (\ref{eq:depolar}), we know that
\begin{equation}
    \pi(e) \sim \begin{pmatrix}d^2\\ \text{supp}(e)\end{pmatrix}p^{\text{supp}(e)}(1-p)^{d^2 - \text{supp}(e)},
\end{equation}
which is a binomial distribution sharply peaked around a maximum $e^\star = \argmax_{\text{supp}(e)\subset e} \pi(e)$. In our model, the distribution is more complicated, but characterised by a mean and a notion of variance as well. To tame the curse of dimensionality, we approximate the sum $\sum_e \pi(e)$ through importance sampling. For the 3D plaquette gauge model, we invoke cold-to-hot \textit{quenched annealing} to probe spin configurations with high Boltzmann weights more effectively, and therefore provide a more accurate approximation of the partition function.

\subsection*{Finite Size Scaling (FSS) for 2D}

For finite system sizes, $m^2$ as defined in Eq. (\ref{eq:mag}) is not a practical order parameter since it encapsulated global properties that are only exact in the thermodynamic limit. This motivates us to turn to a finite size scaling (FSS) framework. Using Eq. (\ref{eq:therm}), we can define the average magnetic susceptibility
\begin{equation}
    \langle\chi(\vec{k})\rangle = \frac{1}{L^2}\biggl\langle\bigg|\sum_j s_j \text{exp}\left(i \vec{k}\cdot \vec{r}_j\right)\bigg|^2\biggl\rangle,
\end{equation}
where $\vec{r}_j$ is the position coordinate vector of the $j$-th spin $s_j$ and momentum vector $\vec{k}$. Following \cite{strongresilience}, we can define the 2-point finite-size correlation function
\begin{equation}
    \xi = \frac{L}{2\pi}\sqrt{\frac{\chi(\vec{0})}{\chi(2\pi/L \hat{n})}-1},
\end{equation}
where $\hat{n}$ is one of the $d$ unit basis vectors in $d$ dimensions. Near the phase transition, finite-size scaling happens according to
\begin{equation}
    \xi = L\cdot \varphi\left(L^{1/\nu}(T-T_c)\right),
\end{equation}
where $\varphi(\cdot)$ is some universal dimensionless function, and $\nu$ is a critical exponent. Simulations were all performed on the GPU to boost efficiency, batching parallel runs by temperature points and error configurations. The plaquette gauge model has local gauge invariance with a local order parameter, so we can not employ FSS.

\subsection*{Wilson loops for (2+1)D}

We construct Wilson loops using square loops along the $xy$-planes of the model, since Pauli twirling shows there is more disorder along spacelike plaquettes than timelike plaquettes. We average over all Wilson loops that can be constructed, because of the isotropy of our model along spacelike plaquettes. In case of erasure, we only add loops to the thermodynamic averaging contribution that have no erased qubits on the loop, only inside. For $r\gtrapprox 0.15 $, this signifies a major breakdown in our simulations. To ensure a good approximation to the partition function, all simulations are a result of 25000 equilibration steps and 25000 Monte Carlo samples.

\bibliographystyle{apsrev4-1}
\bibliography{Bibliography.bib}

\begin{thebibliography}{67}%
\makeatletter
\providecommand \@ifxundefined [1]{%
 \@ifx{#1\undefined}
}%
\providecommand \@ifnum [1]{%
 \ifnum #1\expandafter \@firstoftwo
 \else \expandafter \@secondoftwo
 \fi
}%
\providecommand \@ifx [1]{%
 \ifx #1\expandafter \@firstoftwo
 \else \expandafter \@secondoftwo
 \fi
}%
\providecommand \natexlab [1]{#1}%
\providecommand \enquote  [1]{``#1''}%
\providecommand \bibnamefont  [1]{#1}%
\providecommand \bibfnamefont [1]{#1}%
\providecommand \citenamefont [1]{#1}%
\providecommand \href@noop [0]{\@secondoftwo}%
\providecommand \href [0]{\begingroup \@sanitize@url \@href}%
\providecommand \@href[1]{\@@startlink{#1}\@@href}%
\providecommand \@@href[1]{\endgroup#1\@@endlink}%
\providecommand \@sanitize@url [0]{\catcode `\\12\catcode `\$12\catcode `\&12\catcode `\#12\catcode `\^12\catcode `\_12\catcode `\%12\relax}%
\providecommand \@@startlink[1]{}%
\providecommand \@@endlink[0]{}%
\providecommand \url  [0]{\begingroup\@sanitize@url \@url }%
\providecommand \@url [1]{\endgroup\@href {#1}{\urlprefix }}%
\providecommand \urlprefix  [0]{URL }%
\providecommand \Eprint [0]{\href }%
\providecommand \doibase [0]{http://dx.doi.org/}%
\providecommand \selectlanguage [0]{\@gobble}%
\providecommand \bibinfo  [0]{\@secondoftwo}%
\providecommand \bibfield  [0]{\@secondoftwo}%
\providecommand \translation [1]{[#1]}%
\providecommand \BibitemOpen [0]{}%
\providecommand \bibitemStop [0]{}%
\providecommand \bibitemNoStop [0]{.\EOS\space}%
\providecommand \EOS [0]{\spacefactor3000\relax}%
\providecommand \BibitemShut  [1]{\csname bibitem#1\endcsname}%
\let\auto@bib@innerbib\@empty
\bibitem [{\citenamefont {Preskill}(2018)}]{nisq}%
  \BibitemOpen
  \bibfield  {author} {\bibinfo {author} {\bibfnamefont {J.}~\bibnamefont {Preskill}},\ }\href {\doibase 10.22331/q-2018-08-06-79} {\bibfield  {journal} {\bibinfo  {journal} {Quantum}\ }\textbf {\bibinfo {volume} {2}},\ \bibinfo {pages} {79} (\bibinfo {year} {2018})}\BibitemShut {NoStop}%
\bibitem [{\citenamefont {Terhal}(2015)}]{terhal}%
  \BibitemOpen
  \bibfield  {author} {\bibinfo {author} {\bibfnamefont {B.~M.}\ \bibnamefont {Terhal}},\ }\href {\doibase 10.1103/revmodphys.87.307} {\bibfield  {journal} {\bibinfo  {journal} {Reviews of Modern Physics}\ }\textbf {\bibinfo {volume} {87}},\ \bibinfo {pages} {307} (\bibinfo {year} {2015})}\BibitemShut {NoStop}%
\bibitem [{\citenamefont {Kitaev}(2003)}]{Kitaevtoriccode}%
  \BibitemOpen
  \bibfield  {author} {\bibinfo {author} {\bibfnamefont {A.}~\bibnamefont {Kitaev}},\ }\href {\doibase 10.1016/s0003-4916(02)00018-0} {\bibfield  {journal} {\bibinfo  {journal} {Annals of Physics}\ }\textbf {\bibinfo {volume} {303}},\ \bibinfo {pages} {2} (\bibinfo {year} {2003})}\BibitemShut {NoStop}%
\bibitem [{\citenamefont {Fowler}\ \emph {et~al.}(2012)\citenamefont {Fowler}, \citenamefont {Mariantoni}, \citenamefont {Martinis},\ and\ \citenamefont {Cleland}}]{fowler}%
  \BibitemOpen
  \bibfield  {author} {\bibinfo {author} {\bibfnamefont {A.~G.}\ \bibnamefont {Fowler}}, \bibinfo {author} {\bibfnamefont {M.}~\bibnamefont {Mariantoni}}, \bibinfo {author} {\bibfnamefont {J.~M.}\ \bibnamefont {Martinis}}, \ and\ \bibinfo {author} {\bibfnamefont {A.~N.}\ \bibnamefont {Cleland}},\ }\href {\doibase 10.1103/physreva.86.032324} {\bibfield  {journal} {\bibinfo  {journal} {Physical Review A}\ }\textbf {\bibinfo {volume} {86}} (\bibinfo {year} {2012}),\ 10.1103/physreva.86.032324}\BibitemShut {NoStop}%
\bibitem [{\citenamefont {Litinski}(2019)}]{agameofsurfacecodes}%
  \BibitemOpen
  \bibfield  {author} {\bibinfo {author} {\bibfnamefont {D.}~\bibnamefont {Litinski}},\ }\href {\doibase 10.22331/q-2019-03-05-128} {\bibfield  {journal} {\bibinfo  {journal} {Quantum}\ }\textbf {\bibinfo {volume} {3}},\ \bibinfo {pages} {128} (\bibinfo {year} {2019})}\BibitemShut {NoStop}%
\bibitem [{\citenamefont {{\relax Google Quantum AI}}(2023)}]{googleAI}%
  \BibitemOpen
  \bibfield  {author} {\bibinfo {author} {\bibnamefont {{\relax Google Quantum AI}}},\ }\href@noop {} {\bibfield  {journal} {\bibinfo  {journal} {Nature}\ }\textbf {\bibinfo {volume} {614}},\ \bibinfo {pages} {676} (\bibinfo {year} {2023})}\BibitemShut {NoStop}%
\bibitem [{\citenamefont {{\relax Google Quantum AI}}(2024)}]{googleAI2}%
  \BibitemOpen
  \bibfield  {author} {\bibinfo {author} {\bibnamefont {{\relax Google Quantum AI}}},\ }\href {https://arxiv.org/abs/2408.13687} {\enquote {\bibinfo {title} {Quantum error correction below the surface code threshold},}\ } (\bibinfo {year} {2024}),\ \Eprint {http://arxiv.org/abs/2408.13687} {arXiv:2408.13687 [quant-ph]} \BibitemShut {NoStop}%
\bibitem [{\citenamefont {Jenkins}\ \emph {et~al.}(2022)\citenamefont {Jenkins}, \citenamefont {Lis}, \citenamefont {Senoo}, \citenamefont {McGrew},\ and\ \citenamefont {Kaufman}}]{clock_kaufman}%
  \BibitemOpen
  \bibfield  {author} {\bibinfo {author} {\bibfnamefont {A.}~\bibnamefont {Jenkins}}, \bibinfo {author} {\bibfnamefont {J.~W.}\ \bibnamefont {Lis}}, \bibinfo {author} {\bibfnamefont {A.}~\bibnamefont {Senoo}}, \bibinfo {author} {\bibfnamefont {W.~F.}\ \bibnamefont {McGrew}}, \ and\ \bibinfo {author} {\bibfnamefont {A.~M.}\ \bibnamefont {Kaufman}},\ }\href {\doibase 10.1103/physrevx.12.021027} {\bibfield  {journal} {\bibinfo  {journal} {Physical Review X}\ }\textbf {\bibinfo {volume} {12}} (\bibinfo {year} {2022}),\ 10.1103/physrevx.12.021027}\BibitemShut {NoStop}%
\bibitem [{\citenamefont {Ma}\ \emph {et~al.}(2022)\citenamefont {Ma}, \citenamefont {Burgers}, \citenamefont {Liu}, \citenamefont {Wilson}, \citenamefont {Zhang},\ and\ \citenamefont {Thompson}}]{clock_thompson}%
  \BibitemOpen
  \bibfield  {author} {\bibinfo {author} {\bibfnamefont {S.}~\bibnamefont {Ma}}, \bibinfo {author} {\bibfnamefont {A.~P.}\ \bibnamefont {Burgers}}, \bibinfo {author} {\bibfnamefont {G.}~\bibnamefont {Liu}}, \bibinfo {author} {\bibfnamefont {J.}~\bibnamefont {Wilson}}, \bibinfo {author} {\bibfnamefont {B.}~\bibnamefont {Zhang}}, \ and\ \bibinfo {author} {\bibfnamefont {J.~D.}\ \bibnamefont {Thompson}},\ }\href {\doibase 10.1103/physrevx.12.021028} {\bibfield  {journal} {\bibinfo  {journal} {Physical Review X}\ }\textbf {\bibinfo {volume} {12}} (\bibinfo {year} {2022}),\ 10.1103/physrevx.12.021028}\BibitemShut {NoStop}%
\bibitem [{\citenamefont {Beugnon}\ \emph {et~al.}(2007)\citenamefont {Beugnon}, \citenamefont {Tuchendler}, \citenamefont {Marion}, \citenamefont {Ga{\"e}tan}, \citenamefont {Miroshnychenko}, \citenamefont {Sortais}, \citenamefont {Lance}, \citenamefont {Jones}, \citenamefont {Messin}, \citenamefont {Browaeys},\ and\ \citenamefont {Grangier}}]{eiffel}%
  \BibitemOpen
  \bibfield  {author} {\bibinfo {author} {\bibfnamefont {J.}~\bibnamefont {Beugnon}}, \bibinfo {author} {\bibfnamefont {C.}~\bibnamefont {Tuchendler}}, \bibinfo {author} {\bibfnamefont {H.}~\bibnamefont {Marion}}, \bibinfo {author} {\bibfnamefont {A.}~\bibnamefont {Ga{\"e}tan}}, \bibinfo {author} {\bibfnamefont {Y.}~\bibnamefont {Miroshnychenko}}, \bibinfo {author} {\bibfnamefont {Y.~R.~P.}\ \bibnamefont {Sortais}}, \bibinfo {author} {\bibfnamefont {A.~M.}\ \bibnamefont {Lance}}, \bibinfo {author} {\bibfnamefont {M.~P.~A.}\ \bibnamefont {Jones}}, \bibinfo {author} {\bibfnamefont {G.}~\bibnamefont {Messin}}, \bibinfo {author} {\bibfnamefont {A.}~\bibnamefont {Browaeys}}, \ and\ \bibinfo {author} {\bibfnamefont {P.}~\bibnamefont {Grangier}},\ }\href {\doibase 10.1038/nphys698} {\bibfield  {journal} {\bibinfo  {journal} {Nature Physics}\ }\textbf {\bibinfo {volume} {3}},\ \bibinfo {pages} {696} (\bibinfo {year} {2007})}\BibitemShut {NoStop}%
\bibitem [{\citenamefont {Bluvstein}\ \emph {et~al.}(2022)\citenamefont {Bluvstein}, \citenamefont {Levine}, \citenamefont {Semeghini}, \citenamefont {Wang}, \citenamefont {Ebadi}, \citenamefont {Kalinowski}, \citenamefont {Keesling}, \citenamefont {Maskara}, \citenamefont {Pichler}, \citenamefont {Greiner}, \citenamefont {Vuleti{\'c}},\ and\ \citenamefont {Lukin}}]{movabletweezers}%
  \BibitemOpen
  \bibfield  {author} {\bibinfo {author} {\bibfnamefont {D.}~\bibnamefont {Bluvstein}}, \bibinfo {author} {\bibfnamefont {H.}~\bibnamefont {Levine}}, \bibinfo {author} {\bibfnamefont {G.}~\bibnamefont {Semeghini}}, \bibinfo {author} {\bibfnamefont {T.~T.}\ \bibnamefont {Wang}}, \bibinfo {author} {\bibfnamefont {S.}~\bibnamefont {Ebadi}}, \bibinfo {author} {\bibfnamefont {M.}~\bibnamefont {Kalinowski}}, \bibinfo {author} {\bibfnamefont {A.}~\bibnamefont {Keesling}}, \bibinfo {author} {\bibfnamefont {N.}~\bibnamefont {Maskara}}, \bibinfo {author} {\bibfnamefont {H.}~\bibnamefont {Pichler}}, \bibinfo {author} {\bibfnamefont {M.}~\bibnamefont {Greiner}}, \bibinfo {author} {\bibfnamefont {V.}~\bibnamefont {Vuleti{\'c}}}, \ and\ \bibinfo {author} {\bibfnamefont {M.~D.}\ \bibnamefont {Lukin}},\ }\href {\doibase 10.1038/s41586-022-04592-6} {\bibfield  {journal} {\bibinfo  {journal} {Nature}\ }\textbf {\bibinfo {volume} {604}},\ \bibinfo {pages} {451} (\bibinfo {year} {2022})}\BibitemShut {NoStop}%
\bibitem [{\citenamefont {Bluvstein}\ \emph {et~al.}(2024)\citenamefont {Bluvstein}, \citenamefont {Evered}, \citenamefont {Geim}, \citenamefont {Li}, \citenamefont {Zhou}, \citenamefont {Manovitz}, \citenamefont {Ebadi}, \citenamefont {Cain}, \citenamefont {Kalinowski}, \citenamefont {Hangleiter} \emph {et~al.}}]{lukinQEC}%
  \BibitemOpen
  \bibfield  {author} {\bibinfo {author} {\bibfnamefont {D.}~\bibnamefont {Bluvstein}}, \bibinfo {author} {\bibfnamefont {S.~J.}\ \bibnamefont {Evered}}, \bibinfo {author} {\bibfnamefont {A.~A.}\ \bibnamefont {Geim}}, \bibinfo {author} {\bibfnamefont {S.~H.}\ \bibnamefont {Li}}, \bibinfo {author} {\bibfnamefont {H.}~\bibnamefont {Zhou}}, \bibinfo {author} {\bibfnamefont {T.}~\bibnamefont {Manovitz}}, \bibinfo {author} {\bibfnamefont {S.}~\bibnamefont {Ebadi}}, \bibinfo {author} {\bibfnamefont {M.}~\bibnamefont {Cain}}, \bibinfo {author} {\bibfnamefont {M.}~\bibnamefont {Kalinowski}}, \bibinfo {author} {\bibfnamefont {D.}~\bibnamefont {Hangleiter}},  \emph {et~al.},\ }\href@noop {} {\bibfield  {journal} {\bibinfo  {journal} {Nature}\ }\textbf {\bibinfo {volume} {626}},\ \bibinfo {pages} {58} (\bibinfo {year} {2024})}\BibitemShut {NoStop}%
\bibitem [{\citenamefont {Nishimori}(2001)}]{spinglass}%
  \BibitemOpen
  \bibfield  {author} {\bibinfo {author} {\bibfnamefont {H.}~\bibnamefont {Nishimori}},\ }\href@noop {} {\emph {\bibinfo {title} {Statistical physics of spin glasses and information processing: an introduction}}},\ \bibinfo {number} {111}\ (\bibinfo  {publisher} {Clarendon Press},\ \bibinfo {year} {2001})\BibitemShut {NoStop}%
\bibitem [{\citenamefont {Dennis}\ \emph {et~al.}(2002)\citenamefont {Dennis}, \citenamefont {Kitaev}, \citenamefont {Landahl},\ and\ \citenamefont {Preskill}}]{topologicalquantummemory}%
  \BibitemOpen
  \bibfield  {author} {\bibinfo {author} {\bibfnamefont {E.}~\bibnamefont {Dennis}}, \bibinfo {author} {\bibfnamefont {A.}~\bibnamefont {Kitaev}}, \bibinfo {author} {\bibfnamefont {A.}~\bibnamefont {Landahl}}, \ and\ \bibinfo {author} {\bibfnamefont {J.}~\bibnamefont {Preskill}},\ }\href {\doibase 10.1063/1.1499754} {\bibfield  {journal} {\bibinfo  {journal} {Journal of Mathematical Physics}\ }\textbf {\bibinfo {volume} {43}},\ \bibinfo {pages} {4452} (\bibinfo {year} {2002})}\BibitemShut {NoStop}%
\bibitem [{\citenamefont {Saffman}\ \emph {et~al.}(2010)\citenamefont {Saffman}, \citenamefont {Walker},\ and\ \citenamefont {M{\"o}lmer}}]{saffman}%
  \BibitemOpen
  \bibfield  {author} {\bibinfo {author} {\bibfnamefont {M.}~\bibnamefont {Saffman}}, \bibinfo {author} {\bibfnamefont {T.~G.}\ \bibnamefont {Walker}}, \ and\ \bibinfo {author} {\bibfnamefont {K.}~\bibnamefont {M{\"o}lmer}},\ }\href {\doibase 10.1103/RevModPhys.82.2313} {\bibfield  {journal} {\bibinfo  {journal} {Rev. Mod. Phys.}\ }\textbf {\bibinfo {volume} {82}},\ \bibinfo {pages} {2313} (\bibinfo {year} {2010})}\BibitemShut {NoStop}%
\bibitem [{\citenamefont {Sahay}\ \emph {et~al.}(2023)\citenamefont {Sahay}, \citenamefont {Jin}, \citenamefont {Claes}, \citenamefont {Thompson},\ and\ \citenamefont {Puri}}]{biasederasure}%
  \BibitemOpen
  \bibfield  {author} {\bibinfo {author} {\bibfnamefont {K.}~\bibnamefont {Sahay}}, \bibinfo {author} {\bibfnamefont {J.}~\bibnamefont {Jin}}, \bibinfo {author} {\bibfnamefont {J.}~\bibnamefont {Claes}}, \bibinfo {author} {\bibfnamefont {J.~D.}\ \bibnamefont {Thompson}}, \ and\ \bibinfo {author} {\bibfnamefont {S.}~\bibnamefont {Puri}},\ }\href {\doibase 10.1103/physrevx.13.041013} {\bibfield  {journal} {\bibinfo  {journal} {Physical Review X}\ }\textbf {\bibinfo {volume} {13}} (\bibinfo {year} {2023}),\ 10.1103/physrevx.13.041013}\BibitemShut {NoStop}%
\bibitem [{\citenamefont {Norcia}\ \emph {et~al.}(2018)\citenamefont {Norcia}, \citenamefont {Young},\ and\ \citenamefont {Kaufman}}]{linewidth}%
  \BibitemOpen
  \bibfield  {author} {\bibinfo {author} {\bibfnamefont {M.~A.}\ \bibnamefont {Norcia}}, \bibinfo {author} {\bibfnamefont {A.~W.}\ \bibnamefont {Young}}, \ and\ \bibinfo {author} {\bibfnamefont {A.~M.}\ \bibnamefont {Kaufman}},\ }\href {\doibase 10.1103/PhysRevX.8.041054} {\bibfield  {journal} {\bibinfo  {journal} {Phys. Rev. X}\ }\textbf {\bibinfo {volume} {8}},\ \bibinfo {pages} {041054} (\bibinfo {year} {2018})}\BibitemShut {NoStop}%
\bibitem [{\citenamefont {Evered}\ \emph {et~al.}(2023)\citenamefont {Evered}, \citenamefont {Bluvstein}, \citenamefont {Kalinowski}, \citenamefont {Ebadi}, \citenamefont {Manovitz}, \citenamefont {Zhou}, \citenamefont {Li}, \citenamefont {Geim}, \citenamefont {Wang}, \citenamefont {Maskara}, \citenamefont {Levine}, \citenamefont {Semeghini}, \citenamefont {Greiner}, \citenamefont {Vuleti{\'c}},\ and\ \citenamefont {Lukin}}]{readout}%
  \BibitemOpen
  \bibfield  {author} {\bibinfo {author} {\bibfnamefont {S.~J.}\ \bibnamefont {Evered}}, \bibinfo {author} {\bibfnamefont {D.}~\bibnamefont {Bluvstein}}, \bibinfo {author} {\bibfnamefont {M.}~\bibnamefont {Kalinowski}}, \bibinfo {author} {\bibfnamefont {S.}~\bibnamefont {Ebadi}}, \bibinfo {author} {\bibfnamefont {T.}~\bibnamefont {Manovitz}}, \bibinfo {author} {\bibfnamefont {H.}~\bibnamefont {Zhou}}, \bibinfo {author} {\bibfnamefont {S.~H.}\ \bibnamefont {Li}}, \bibinfo {author} {\bibfnamefont {A.~A.}\ \bibnamefont {Geim}}, \bibinfo {author} {\bibfnamefont {T.~T.}\ \bibnamefont {Wang}}, \bibinfo {author} {\bibfnamefont {N.}~\bibnamefont {Maskara}}, \bibinfo {author} {\bibfnamefont {H.}~\bibnamefont {Levine}}, \bibinfo {author} {\bibfnamefont {G.}~\bibnamefont {Semeghini}}, \bibinfo {author} {\bibfnamefont {M.}~\bibnamefont {Greiner}}, \bibinfo {author} {\bibfnamefont {V.}~\bibnamefont {Vuleti{\'c}}}, \ and\ \bibinfo {author} {\bibfnamefont {M.~D.}\ \bibnamefont {Lukin}},\ }\href {\doibase
  10.1038/s41586-023-06481-y} {\bibfield  {journal} {\bibinfo  {journal} {Nature}\ }\textbf {\bibinfo {volume} {622}},\ \bibinfo {pages} {268} (\bibinfo {year} {2023})}\BibitemShut {NoStop}%
\bibitem [{\citenamefont {Xu}\ \emph {et~al.}(2023)\citenamefont {Xu}, \citenamefont {Bonilla~Ataides}, \citenamefont {Pattison}, \citenamefont {Raveendran}, \citenamefont {Bluvstein}, \citenamefont {Wurtz}, \citenamefont {Vasic}, \citenamefont {Lukin}, \citenamefont {Jiang},\ and\ \citenamefont {Zhou}}]{lukinLDPC}%
  \BibitemOpen
  \bibfield  {author} {\bibinfo {author} {\bibfnamefont {Q.}~\bibnamefont {Xu}}, \bibinfo {author} {\bibfnamefont {J.~P.}\ \bibnamefont {Bonilla~Ataides}}, \bibinfo {author} {\bibfnamefont {C.}~\bibnamefont {Pattison}}, \bibinfo {author} {\bibfnamefont {N.}~\bibnamefont {Raveendran}}, \bibinfo {author} {\bibfnamefont {D.}~\bibnamefont {Bluvstein}}, \bibinfo {author} {\bibfnamefont {J.}~\bibnamefont {Wurtz}}, \bibinfo {author} {\bibfnamefont {B.}~\bibnamefont {Vasic}}, \bibinfo {author} {\bibfnamefont {M.}~\bibnamefont {Lukin}}, \bibinfo {author} {\bibfnamefont {L.}~\bibnamefont {Jiang}}, \ and\ \bibinfo {author} {\bibfnamefont {H.}~\bibnamefont {Zhou}},\ }\href@noop {} {\bibfield  {journal} {\bibinfo  {journal} {Nature}\ } (\bibinfo {year} {2023})}\BibitemShut {NoStop}%
\bibitem [{\citenamefont {Bravyi}\ \emph {et~al.}(2023)\citenamefont {Bravyi}, \citenamefont {Cross}, \citenamefont {Gambetta}, \citenamefont {Maslov}, \citenamefont {Rall},\ and\ \citenamefont {Yoder}}]{ibmLDPC}%
  \BibitemOpen
  \bibfield  {author} {\bibinfo {author} {\bibfnamefont {S.}~\bibnamefont {Bravyi}}, \bibinfo {author} {\bibfnamefont {A.}~\bibnamefont {Cross}}, \bibinfo {author} {\bibfnamefont {J.}~\bibnamefont {Gambetta}}, \bibinfo {author} {\bibfnamefont {D.}~\bibnamefont {Maslov}}, \bibinfo {author} {\bibfnamefont {P.}~\bibnamefont {Rall}}, \ and\ \bibinfo {author} {\bibfnamefont {T.}~\bibnamefont {Yoder}},\ }\href@noop {} {\bibfield  {journal} {\bibinfo  {journal} {Nature}\ } (\bibinfo {year} {2023})}\BibitemShut {NoStop}%
\bibitem [{\citenamefont {Geller}\ and\ \citenamefont {Zhou}(2013)}]{twirling}%
  \BibitemOpen
  \bibfield  {author} {\bibinfo {author} {\bibfnamefont {M.~R.}\ \bibnamefont {Geller}}\ and\ \bibinfo {author} {\bibfnamefont {Z.}~\bibnamefont {Zhou}},\ }\href {\doibase 10.1103/physreva.88.012314} {\bibfield  {journal} {\bibinfo  {journal} {Physical Review A}\ }\textbf {\bibinfo {volume} {88}} (\bibinfo {year} {2013}),\ 10.1103/physreva.88.012314}\BibitemShut {NoStop}%
\bibitem [{\citenamefont {Dankert}\ \emph {et~al.}(2009)\citenamefont {Dankert}, \citenamefont {Cleve}, \citenamefont {Emerson},\ and\ \citenamefont {Livine}}]{twirling2}%
  \BibitemOpen
  \bibfield  {author} {\bibinfo {author} {\bibfnamefont {C.}~\bibnamefont {Dankert}}, \bibinfo {author} {\bibfnamefont {R.}~\bibnamefont {Cleve}}, \bibinfo {author} {\bibfnamefont {J.}~\bibnamefont {Emerson}}, \ and\ \bibinfo {author} {\bibfnamefont {E.}~\bibnamefont {Livine}},\ }\href {\doibase 10.1103/PhysRevA.80.012304} {\bibfield  {journal} {\bibinfo  {journal} {Phys. Rev. A}\ }\textbf {\bibinfo {volume} {80}},\ \bibinfo {pages} {012304} (\bibinfo {year} {2009})}\BibitemShut {NoStop}%
\bibitem [{\citenamefont {Jandura}\ and\ \citenamefont {Pupillo}(2024)}]{sventwirl}%
  \BibitemOpen
  \bibfield  {author} {\bibinfo {author} {\bibfnamefont {S.}~\bibnamefont {Jandura}}\ and\ \bibinfo {author} {\bibfnamefont {G.}~\bibnamefont {Pupillo}},\ }\href@noop {} {\enquote {\bibinfo {title} {Surface code stabilizer measurements for {R}ydberg atoms},}\ } (\bibinfo {year} {2024}),\ \Eprint {http://arxiv.org/abs/2405.16621} {arXiv:2405.16621} \BibitemShut {NoStop}%
\bibitem [{\citenamefont {Poole}\ \emph {et~al.}(2024)\citenamefont {Poole}, \citenamefont {Graham}, \citenamefont {Perlin}, \citenamefont {Otten},\ and\ \citenamefont {Saffman}}]{saffmantrans}%
  \BibitemOpen
  \bibfield  {author} {\bibinfo {author} {\bibfnamefont {C.}~\bibnamefont {Poole}}, \bibinfo {author} {\bibfnamefont {T.~M.}\ \bibnamefont {Graham}}, \bibinfo {author} {\bibfnamefont {M.~A.}\ \bibnamefont {Perlin}}, \bibinfo {author} {\bibfnamefont {M.}~\bibnamefont {Otten}}, \ and\ \bibinfo {author} {\bibfnamefont {M.}~\bibnamefont {Saffman}},\ }\href@noop {} {\enquote {\bibinfo {title} {Architecture for fast implementation of qldpc codes with optimized rydberg gates},}\ } (\bibinfo {year} {2024}),\ \Eprint {http://arxiv.org/abs/2404.18809} {arXiv:2404.18809} \BibitemShut {NoStop}%
\bibitem [{\citenamefont {Scholl}\ \emph {et~al.}(2023{\natexlab{a}})\citenamefont {Scholl}, \citenamefont {Shaw}, \citenamefont {Tsai}, \citenamefont {Finkelstein}, \citenamefont {Choi},\ and\ \citenamefont {Endres}}]{98erasure}%
  \BibitemOpen
  \bibfield  {author} {\bibinfo {author} {\bibfnamefont {P.}~\bibnamefont {Scholl}}, \bibinfo {author} {\bibfnamefont {A.~L.}\ \bibnamefont {Shaw}}, \bibinfo {author} {\bibfnamefont {R.~B.-S.}\ \bibnamefont {Tsai}}, \bibinfo {author} {\bibfnamefont {R.}~\bibnamefont {Finkelstein}}, \bibinfo {author} {\bibfnamefont {J.}~\bibnamefont {Choi}}, \ and\ \bibinfo {author} {\bibfnamefont {M.}~\bibnamefont {Endres}},\ }\href {\doibase 10.1038/s41586-023-06516-4} {\bibfield  {journal} {\bibinfo  {journal} {Nature}\ }\textbf {\bibinfo {volume} {622}},\ \bibinfo {pages} {273} (\bibinfo {year} {2023}{\natexlab{a}})}\BibitemShut {NoStop}%
\bibitem [{\citenamefont {Cover}\ and\ \citenamefont {Thomas}(2006)}]{infotheory}%
  \BibitemOpen
  \bibfield  {author} {\bibinfo {author} {\bibfnamefont {T.~M.}\ \bibnamefont {Cover}}\ and\ \bibinfo {author} {\bibfnamefont {J.~A.}\ \bibnamefont {Thomas}},\ }\href@noop {} {\emph {\bibinfo {title} {Elements of Information Theory (Wiley Series in Telecommunications and Signal Processing)}}}\ (\bibinfo  {publisher} {Wiley-Interscience},\ \bibinfo {address} {USA},\ \bibinfo {year} {2006})\BibitemShut {NoStop}%
\bibitem [{\citenamefont {Cong}\ \emph {et~al.}(2022)\citenamefont {Cong}, \citenamefont {Levine}, \citenamefont {Keesling}, \citenamefont {Bluvstein}, \citenamefont {Wang},\ and\ \citenamefont {Lukin}}]{iriscong}%
  \BibitemOpen
  \bibfield  {author} {\bibinfo {author} {\bibfnamefont {I.}~\bibnamefont {Cong}}, \bibinfo {author} {\bibfnamefont {H.}~\bibnamefont {Levine}}, \bibinfo {author} {\bibfnamefont {A.}~\bibnamefont {Keesling}}, \bibinfo {author} {\bibfnamefont {D.}~\bibnamefont {Bluvstein}}, \bibinfo {author} {\bibfnamefont {S.-T.}\ \bibnamefont {Wang}}, \ and\ \bibinfo {author} {\bibfnamefont {M.~D.}\ \bibnamefont {Lukin}},\ }\href {\doibase 10.1103/PhysRevX.12.021049} {\bibfield  {journal} {\bibinfo  {journal} {Phys. Rev. X}\ }\textbf {\bibinfo {volume} {12}},\ \bibinfo {pages} {021049} (\bibinfo {year} {2022})}\BibitemShut {NoStop}%
\bibitem [{\citenamefont {Calderbank}\ and\ \citenamefont {Shor}(1996)}]{CS}%
  \BibitemOpen
  \bibfield  {author} {\bibinfo {author} {\bibfnamefont {A.~R.}\ \bibnamefont {Calderbank}}\ and\ \bibinfo {author} {\bibfnamefont {P.~W.}\ \bibnamefont {Shor}},\ }\href {\doibase 10.1103/physreva.54.1098} {\bibfield  {journal} {\bibinfo  {journal} {Physical Review A}\ }\textbf {\bibinfo {volume} {54}},\ \bibinfo {pages} {1098} (\bibinfo {year} {1996})}\BibitemShut {NoStop}%
\bibitem [{\citenamefont {Steane}(1996)}]{S}%
  \BibitemOpen
  \bibfield  {author} {\bibinfo {author} {\bibfnamefont {A.}~\bibnamefont {Steane}},\ }\href {\doibase 10.1098/rspa.1996.0136} {\bibfield  {journal} {\bibinfo  {journal} {Proceedings of the Royal Society of London. Series A: Mathematical, Physical and Engineering Sciences}\ }\textbf {\bibinfo {volume} {452}},\ \bibinfo {pages} {2551} (\bibinfo {year} {1996})}\BibitemShut {NoStop}%
\bibitem [{Note1()}]{Note1}%
  \BibitemOpen
  \bibinfo {note} {$\protect \mathbb {F}_q$ is the finite field of $q$ elements, also called the Galois field, where $q=p^s$ is an integer power of some prime number $p$. $\protect \mathbb {F}_q^n$ is equivalent to the $n$-dimensional field $\protect \mathbb {F}_q\times \protect \cdots \times \protect \mathbb {F}_q$.}\BibitemShut {Stop}%
\bibitem [{Note2()}]{Note2}%
  \BibitemOpen
  \bibinfo {note} {Note that in this Article, we use a depiction of the $d=3$ rotated surface code to represent \protect \textit {any} topological CSS codes, because of its simplicity.}\BibitemShut {Stop}%
\bibitem [{\citenamefont {Jaksch}\ \emph {et~al.}(2000)\citenamefont {Jaksch}, \citenamefont {Cirac}, \citenamefont {Zoller}, \citenamefont {Rolston}, \citenamefont {C{\^ot\'e}},\ and\ \citenamefont {Lukin}}]{jaksch}%
  \BibitemOpen
  \bibfield  {author} {\bibinfo {author} {\bibfnamefont {D.}~\bibnamefont {Jaksch}}, \bibinfo {author} {\bibfnamefont {J.}~\bibnamefont {Cirac}}, \bibinfo {author} {\bibfnamefont {P.}~\bibnamefont {Zoller}}, \bibinfo {author} {\bibfnamefont {S.~L.}\ \bibnamefont {Rolston}}, \bibinfo {author} {\bibfnamefont {R.}~\bibnamefont {C{\^ot\'e}}}, \ and\ \bibinfo {author} {\bibfnamefont {M.}~\bibnamefont {Lukin}},\ }\href {\doibase 10.1103/PhysRevLett.85.2208} {\bibfield  {journal} {\bibinfo  {journal} {Phys. Rev. Lett.}\ }\textbf {\bibinfo {volume} {85}},\ \bibinfo {pages} {2208} (\bibinfo {year} {2000})}\BibitemShut {NoStop}%
\bibitem [{\citenamefont {Jandura}\ and\ \citenamefont {Pupillo}(2022)}]{sven}%
  \BibitemOpen
  \bibfield  {author} {\bibinfo {author} {\bibfnamefont {S.}~\bibnamefont {Jandura}}\ and\ \bibinfo {author} {\bibfnamefont {G.}~\bibnamefont {Pupillo}},\ }\href {https://api.semanticscholar.org/CorpusID:246473360} {\bibfield  {journal} {\bibinfo  {journal} {Quantum}\ }\textbf {\bibinfo {volume} {6}},\ \bibinfo {pages} {712} (\bibinfo {year} {2022})}\BibitemShut {NoStop}%
\bibitem [{\citenamefont {Pagano}\ \emph {et~al.}(2022)\citenamefont {Pagano}, \citenamefont {Weber}, \citenamefont {Jaschke}, \citenamefont {Pfau}, \citenamefont {Meinert}, \citenamefont {Montangero},\ and\ \citenamefont {B\"uchler}}]{alice}%
  \BibitemOpen
  \bibfield  {author} {\bibinfo {author} {\bibfnamefont {A.}~\bibnamefont {Pagano}}, \bibinfo {author} {\bibfnamefont {S.}~\bibnamefont {Weber}}, \bibinfo {author} {\bibfnamefont {D.}~\bibnamefont {Jaschke}}, \bibinfo {author} {\bibfnamefont {T.}~\bibnamefont {Pfau}}, \bibinfo {author} {\bibfnamefont {F.}~\bibnamefont {Meinert}}, \bibinfo {author} {\bibfnamefont {S.}~\bibnamefont {Montangero}}, \ and\ \bibinfo {author} {\bibfnamefont {H.~P.}\ \bibnamefont {B\"uchler}},\ }\href {\doibase 10.1103/PhysRevResearch.4.033019} {\bibfield  {journal} {\bibinfo  {journal} {Phys. Rev. Res.}\ }\textbf {\bibinfo {volume} {4}},\ \bibinfo {pages} {033019} (\bibinfo {year} {2022})}\BibitemShut {NoStop}%
\bibitem [{\citenamefont {Levine}\ \emph {et~al.}(2019)\citenamefont {Levine}, \citenamefont {Keesling}, \citenamefont {Semeghini}, \citenamefont {Omran}, \citenamefont {Wang}, \citenamefont {Ebadi}, \citenamefont {Bernien}, \citenamefont {Greiner}, \citenamefont {Vuleti\ifmmode~\acute{c}\else \'{c}\fi{}}, \citenamefont {Pichler},\ and\ \citenamefont {Lukin}}]{levine}%
  \BibitemOpen
  \bibfield  {author} {\bibinfo {author} {\bibfnamefont {H.}~\bibnamefont {Levine}}, \bibinfo {author} {\bibfnamefont {A.}~\bibnamefont {Keesling}}, \bibinfo {author} {\bibfnamefont {G.}~\bibnamefont {Semeghini}}, \bibinfo {author} {\bibfnamefont {A.}~\bibnamefont {Omran}}, \bibinfo {author} {\bibfnamefont {T.~T.}\ \bibnamefont {Wang}}, \bibinfo {author} {\bibfnamefont {S.}~\bibnamefont {Ebadi}}, \bibinfo {author} {\bibfnamefont {H.}~\bibnamefont {Bernien}}, \bibinfo {author} {\bibfnamefont {M.}~\bibnamefont {Greiner}}, \bibinfo {author} {\bibfnamefont {V.}~\bibnamefont {Vuleti\ifmmode~\acute{c}\else \'{c}\fi{}}}, \bibinfo {author} {\bibfnamefont {H.}~\bibnamefont {Pichler}}, \ and\ \bibinfo {author} {\bibfnamefont {M.~D.}\ \bibnamefont {Lukin}},\ }\href {\doibase 10.1103/PhysRevLett.123.170503} {\bibfield  {journal} {\bibinfo  {journal} {Phys. Rev. Lett.}\ }\textbf {\bibinfo {volume} {123}},\ \bibinfo {pages} {170503} (\bibinfo {year} {2019})}\BibitemShut {NoStop}%
\bibitem [{\citenamefont {Mohan}\ \emph {et~al.}(2023)\citenamefont {Mohan}, \citenamefont {de~Keijzer},\ and\ \citenamefont {Kokkelmans}}]{madhav}%
  \BibitemOpen
  \bibfield  {author} {\bibinfo {author} {\bibfnamefont {M.}~\bibnamefont {Mohan}}, \bibinfo {author} {\bibfnamefont {R.}~\bibnamefont {de~Keijzer}}, \ and\ \bibinfo {author} {\bibfnamefont {S.}~\bibnamefont {Kokkelmans}},\ }\href@noop {} {\bibfield  {journal} {\bibinfo  {journal} {Physical Review Research}\ }\textbf {\bibinfo {volume} {5}},\ \bibinfo {pages} {033052} (\bibinfo {year} {2023})}\BibitemShut {NoStop}%
\bibitem [{\citenamefont {Tian}\ \emph {et~al.}(2023)\citenamefont {Tian}, \citenamefont {Wee}, \citenamefont {Qu}, \citenamefont {Lim}, \citenamefont {Datla}, \citenamefont {Koh},\ and\ \citenamefont {Loh}}]{60microns}%
  \BibitemOpen
  \bibfield  {author} {\bibinfo {author} {\bibfnamefont {W.}~\bibnamefont {Tian}}, \bibinfo {author} {\bibfnamefont {W.~J.}\ \bibnamefont {Wee}}, \bibinfo {author} {\bibfnamefont {A.}~\bibnamefont {Qu}}, \bibinfo {author} {\bibfnamefont {B.~J.~M.}\ \bibnamefont {Lim}}, \bibinfo {author} {\bibfnamefont {P.~R.}\ \bibnamefont {Datla}}, \bibinfo {author} {\bibfnamefont {V.~P.~W.}\ \bibnamefont {Koh}}, \ and\ \bibinfo {author} {\bibfnamefont {H.}~\bibnamefont {Loh}},\ }\href {\doibase 10.1103/physrevapplied.19.034048} {\bibfield  {journal} {\bibinfo  {journal} {Physical Review Applied}\ }\textbf {\bibinfo {volume} {19}} (\bibinfo {year} {2023}),\ 10.1103/physrevapplied.19.034048}\BibitemShut {NoStop}%
\bibitem [{\citenamefont {Covey}\ \emph {et~al.}(2019)\citenamefont {Covey}, \citenamefont {Madjarov}, \citenamefont {Cooper},\ and\ \citenamefont {Endres}}]{meas}%
  \BibitemOpen
  \bibfield  {author} {\bibinfo {author} {\bibfnamefont {J.~P.}\ \bibnamefont {Covey}}, \bibinfo {author} {\bibfnamefont {I.~S.}\ \bibnamefont {Madjarov}}, \bibinfo {author} {\bibfnamefont {A.}~\bibnamefont {Cooper}}, \ and\ \bibinfo {author} {\bibfnamefont {M.}~\bibnamefont {Endres}},\ }\href {\doibase 10.1103/physrevlett.122.173201} {\bibfield  {journal} {\bibinfo  {journal} {Physical Review Letters}\ }\textbf {\bibinfo {volume} {122}} (\bibinfo {year} {2019}),\ 10.1103/physrevlett.122.173201}\BibitemShut {NoStop}%
\bibitem [{\citenamefont {Tan}\ \emph {et~al.}(2024)\citenamefont {Tan}, \citenamefont {Bluvstein}, \citenamefont {Lukin},\ and\ \citenamefont {Cong}}]{transpilationishard}%
  \BibitemOpen
  \bibfield  {author} {\bibinfo {author} {\bibfnamefont {D.~B.}\ \bibnamefont {Tan}}, \bibinfo {author} {\bibfnamefont {D.}~\bibnamefont {Bluvstein}}, \bibinfo {author} {\bibfnamefont {M.~D.}\ \bibnamefont {Lukin}}, \ and\ \bibinfo {author} {\bibfnamefont {J.}~\bibnamefont {Cong}},\ }\href {\doibase 10.22331/q-2024-03-14-1281} {\bibfield  {journal} {\bibinfo  {journal} {Quantum}\ }\textbf {\bibinfo {volume} {8}},\ \bibinfo {pages} {1281} (\bibinfo {year} {2024})}\BibitemShut {NoStop}%
\bibitem [{\citenamefont {Chubb}\ and\ \citenamefont {Flammia}(2021)}]{chubb}%
  \BibitemOpen
  \bibfield  {author} {\bibinfo {author} {\bibfnamefont {C.~T.}\ \bibnamefont {Chubb}}\ and\ \bibinfo {author} {\bibfnamefont {S.~T.}\ \bibnamefont {Flammia}},\ }\href {\doibase 10.4171/aihpd/105} {\bibfield  {journal} {\bibinfo  {journal} {Annales de l'Institut Henri Poincar{\'e} D}\ }\textbf {\bibinfo {volume} {8}},\ \bibinfo {pages} {269} (\bibinfo {year} {2021})}\BibitemShut {NoStop}%
\bibitem [{\citenamefont {Wang}\ \emph {et~al.}(2003)\citenamefont {Wang}, \citenamefont {Harrington},\ and\ \citenamefont {Preskill}}]{higgs}%
  \BibitemOpen
  \bibfield  {author} {\bibinfo {author} {\bibfnamefont {C.}~\bibnamefont {Wang}}, \bibinfo {author} {\bibfnamefont {J.}~\bibnamefont {Harrington}}, \ and\ \bibinfo {author} {\bibfnamefont {J.}~\bibnamefont {Preskill}},\ }\href {\doibase 10.1016/s0003-4916(02)00019-2} {\bibfield  {journal} {\bibinfo  {journal} {Annals of Physics}\ }\textbf {\bibinfo {volume} {303}},\ \bibinfo {pages} {31} (\bibinfo {year} {2003})}\BibitemShut {NoStop}%
\bibitem [{\citenamefont {Vodola}\ \emph {et~al.}(2022)\citenamefont {Vodola}, \citenamefont {Rispler}, \citenamefont {Kim},\ and\ \citenamefont {M{\"u}ller}}]{ookstat}%
  \BibitemOpen
  \bibfield  {author} {\bibinfo {author} {\bibfnamefont {D.}~\bibnamefont {Vodola}}, \bibinfo {author} {\bibfnamefont {M.}~\bibnamefont {Rispler}}, \bibinfo {author} {\bibfnamefont {S.}~\bibnamefont {Kim}}, \ and\ \bibinfo {author} {\bibfnamefont {M.}~\bibnamefont {M{\"u}ller}},\ }\href {\doibase 10.22331/q-2022-01-05-618} {\bibfield  {journal} {\bibinfo  {journal} {Quantum}\ }\textbf {\bibinfo {volume} {6}},\ \bibinfo {pages} {618} (\bibinfo {year} {2022})}\BibitemShut {NoStop}%
\bibitem [{\citenamefont {Edmonds}(1965)}]{blossom}%
  \BibitemOpen
  \bibfield  {author} {\bibinfo {author} {\bibfnamefont {J.}~\bibnamefont {Edmonds}},\ }\href {\doibase 10.4153/CJM-1965-045-4} {\bibfield  {journal} {\bibinfo  {journal} {Canadian Journal of Mathematics}\ }\textbf {\bibinfo {volume} {17}},\ \bibinfo {pages} {449} (\bibinfo {year} {1965})}\BibitemShut {NoStop}%
\bibitem [{\citenamefont {Roffe}\ \emph {et~al.}(2020)\citenamefont {Roffe}, \citenamefont {White}, \citenamefont {Burton},\ and\ \citenamefont {Campbell}}]{belief}%
  \BibitemOpen
  \bibfield  {author} {\bibinfo {author} {\bibfnamefont {J.}~\bibnamefont {Roffe}}, \bibinfo {author} {\bibfnamefont {D.~R.}\ \bibnamefont {White}}, \bibinfo {author} {\bibfnamefont {S.}~\bibnamefont {Burton}}, \ and\ \bibinfo {author} {\bibfnamefont {E.}~\bibnamefont {Campbell}},\ }\href {\doibase 10.1103/PhysRevResearch.2.043423} {\bibfield  {journal} {\bibinfo  {journal} {Phys. Rev. Res.}\ }\textbf {\bibinfo {volume} {2}},\ \bibinfo {pages} {043423} (\bibinfo {year} {2020})}\BibitemShut {NoStop}%
\bibitem [{Note3()}]{Note3}%
  \BibitemOpen
  \bibinfo {note} {The signum function $\eta $ is a function that tells us the sign of a quantity, such as a +1 coupling strength for no error, 0 for an erasure, and -1 for a Pauli error.}\BibitemShut {Stop}%
\bibitem [{\citenamefont {Honecker}\ \emph {et~al.}(2001)\citenamefont {Honecker}, \citenamefont {Picco},\ and\ \citenamefont {Pujol}}]{nishimori}%
  \BibitemOpen
  \bibfield  {author} {\bibinfo {author} {\bibfnamefont {A.}~\bibnamefont {Honecker}}, \bibinfo {author} {\bibfnamefont {M.}~\bibnamefont {Picco}}, \ and\ \bibinfo {author} {\bibfnamefont {P.}~\bibnamefont {Pujol}},\ }\href {\doibase 10.1103/PhysRevLett.87.047201} {\bibfield  {journal} {\bibinfo  {journal} {Phys. Rev. Lett.}\ }\textbf {\bibinfo {volume} {87}},\ \bibinfo {pages} {047201} (\bibinfo {year} {2001})},\ \Eprint {http://arxiv.org/abs/cond-mat/0010143} {arXiv:cond-mat/0010143} \BibitemShut {NoStop}%
\bibitem [{\citenamefont {Le~Doussal}\ and\ \citenamefont {Harris}(1988)}]{fixedpoint}%
  \BibitemOpen
  \bibfield  {author} {\bibinfo {author} {\bibfnamefont {P.}~\bibnamefont {Le~Doussal}}\ and\ \bibinfo {author} {\bibfnamefont {A.~B.}\ \bibnamefont {Harris}},\ }\href {\doibase 10.1103/PhysRevLett.61.625} {\bibfield  {journal} {\bibinfo  {journal} {Phys. Rev. Lett.}\ }\textbf {\bibinfo {volume} {61}},\ \bibinfo {pages} {625} (\bibinfo {year} {1988})}\BibitemShut {NoStop}%
\bibitem [{Note4()}]{Note4}%
  \BibitemOpen
  \bibinfo {note} {Note that we dropped the global phase $i$ here, which yields $Y^\dagger = -Y$.}\BibitemShut {Stop}%
\bibitem [{\citenamefont {Reger}\ and\ \citenamefont {Zippelius}(1986)}]{3drbim}%
  \BibitemOpen
  \bibfield  {author} {\bibinfo {author} {\bibfnamefont {J.~D.}\ \bibnamefont {Reger}}\ and\ \bibinfo {author} {\bibfnamefont {A.}~\bibnamefont {Zippelius}},\ }\href {\doibase 10.1103/PhysRevLett.57.3225} {\bibfield  {journal} {\bibinfo  {journal} {Phys. Rev. Lett.}\ }\textbf {\bibinfo {volume} {57}},\ \bibinfo {pages} {3225} (\bibinfo {year} {1986})}\BibitemShut {NoStop}%
\bibitem [{\citenamefont {Gidney}(2021)}]{stim}%
  \BibitemOpen
  \bibfield  {author} {\bibinfo {author} {\bibfnamefont {C.}~\bibnamefont {Gidney}},\ }\href {\doibase 10.22331/q-2021-07-06-497} {\bibfield  {journal} {\bibinfo  {journal} {{Quantum}}\ }\textbf {\bibinfo {volume} {5}},\ \bibinfo {pages} {497} (\bibinfo {year} {2021})}\BibitemShut {NoStop}%
\bibitem [{\citenamefont {Kogut}(1979)}]{kogut}%
  \BibitemOpen
  \bibfield  {author} {\bibinfo {author} {\bibfnamefont {J.~B.}\ \bibnamefont {Kogut}},\ }\href {\doibase 10.1103/RevModPhys.51.659} {\bibfield  {journal} {\bibinfo  {journal} {Rev. Mod. Phys.}\ }\textbf {\bibinfo {volume} {51}},\ \bibinfo {pages} {659} (\bibinfo {year} {1979})}\BibitemShut {NoStop}%
\bibitem [{\citenamefont {Elitzur}(1975)}]{elitzur}%
  \BibitemOpen
  \bibfield  {author} {\bibinfo {author} {\bibfnamefont {S.}~\bibnamefont {Elitzur}},\ }\href {\doibase 10.1103/PhysRevD.12.3978} {\bibfield  {journal} {\bibinfo  {journal} {Phys. Rev. D}\ }\textbf {\bibinfo {volume} {12}},\ \bibinfo {pages} {3978} (\bibinfo {year} {1975})}\BibitemShut {NoStop}%
\bibitem [{\citenamefont {Makeenko}(2002)}]{gauge}%
  \BibitemOpen
  \bibfield  {author} {\bibinfo {author} {\bibfnamefont {Y.}~\bibnamefont {Makeenko}},\ }\enquote {\bibinfo {title} {Gauge fields on a lattice},}\ in\ \href@noop {} {\emph {\bibinfo {booktitle} {Methods of Contemporary Gauge Theory}}},\ \bibinfo {series and number} {Cambridge Monographs on Mathematical Physics}\ (\bibinfo  {publisher} {Cambridge University Press},\ \bibinfo {year} {2002})\ p.~\bibinfo {pages} {99}\BibitemShut {NoStop}%
\bibitem [{\citenamefont {Lee}\ \emph {et~al.}(2022)\citenamefont {Lee}, \citenamefont {Ji}, \citenamefont {Bi},\ and\ \citenamefont {Fisher}}]{thatonephasediagrampaper}%
  \BibitemOpen
  \bibfield  {author} {\bibinfo {author} {\bibfnamefont {J.~Y.}\ \bibnamefont {Lee}}, \bibinfo {author} {\bibfnamefont {W.}~\bibnamefont {Ji}}, \bibinfo {author} {\bibfnamefont {Z.}~\bibnamefont {Bi}}, \ and\ \bibinfo {author} {\bibfnamefont {M.~P.~A.}\ \bibnamefont {Fisher}},\ }\href@noop {} {\enquote {\bibinfo {title} {Decoding measurement-prepared quantum phases and transitions: from ising model to gauge theory, and beyond},}\ } (\bibinfo {year} {2022}),\ \Eprint {http://arxiv.org/abs/2208.11699} {arXiv:2208.11699 [cond-mat.str-el]} \BibitemShut {NoStop}%
\bibitem [{\citenamefont {Suzuki}(1972)}]{fukinuke}%
  \BibitemOpen
  \bibfield  {author} {\bibinfo {author} {\bibfnamefont {M.}~\bibnamefont {Suzuki}},\ }\href {\doibase 10.1103/PhysRevLett.28.507} {\bibfield  {journal} {\bibinfo  {journal} {Phys. Rev. Lett.}\ }\textbf {\bibinfo {volume} {28}},\ \bibinfo {pages} {507} (\bibinfo {year} {1972})}\BibitemShut {NoStop}%
\bibitem [{\citenamefont {Jacobsen}(2014)}]{perclist}%
  \BibitemOpen
  \bibfield  {author} {\bibinfo {author} {\bibfnamefont {J.}~\bibnamefont {Jacobsen}},\ }\href {\doibase 10.1088/1751-8113/47/13/135001} {\bibfield  {journal} {\bibinfo  {journal} {Journal of Physics A: Mathematical and Theoretical}\ }\textbf {\bibinfo {volume} {47}} (\bibinfo {year} {2014}),\ 10.1088/1751-8113/47/13/135001}\BibitemShut {NoStop}%
\bibitem [{\citenamefont {Kesten}(1980)}]{1/2percolation}%
  \BibitemOpen
  \bibfield  {author} {\bibinfo {author} {\bibfnamefont {H.}~\bibnamefont {Kesten}},\ }\href {https://api.semanticscholar.org/CorpusID:3143683} {\bibfield  {journal} {\bibinfo  {journal} {Communications in Mathematical Physics}\ }\textbf {\bibinfo {volume} {74}},\ \bibinfo {pages} {41} (\bibinfo {year} {1980})}\BibitemShut {NoStop}%
\bibitem [{\citenamefont {Caracciolo}\ \emph {et~al.}(1995)\citenamefont {Caracciolo}, \citenamefont {Edwards}, \citenamefont {Ferreira}, \citenamefont {Pelissetto},\ and\ \citenamefont {Sokal}}]{FSS}%
  \BibitemOpen
  \bibfield  {author} {\bibinfo {author} {\bibfnamefont {S.}~\bibnamefont {Caracciolo}}, \bibinfo {author} {\bibfnamefont {R.~G.}\ \bibnamefont {Edwards}}, \bibinfo {author} {\bibfnamefont {S.~J.}\ \bibnamefont {Ferreira}}, \bibinfo {author} {\bibfnamefont {A.}~\bibnamefont {Pelissetto}}, \ and\ \bibinfo {author} {\bibfnamefont {A.~D.}\ \bibnamefont {Sokal}},\ }\href {\doibase 10.1103/PhysRevLett.74.2969} {\bibfield  {journal} {\bibinfo  {journal} {Phys. Rev. Lett.}\ }\textbf {\bibinfo {volume} {74}},\ \bibinfo {pages} {2969} (\bibinfo {year} {1995})}\BibitemShut {NoStop}%
\bibitem [{\citenamefont {Palassini}\ and\ \citenamefont {Caracciolo}(1999)}]{FSS2}%
  \BibitemOpen
  \bibfield  {author} {\bibinfo {author} {\bibfnamefont {M.}~\bibnamefont {Palassini}}\ and\ \bibinfo {author} {\bibfnamefont {S.}~\bibnamefont {Caracciolo}},\ }\href {\doibase 10.1103/PhysRevLett.82.5128} {\bibfield  {journal} {\bibinfo  {journal} {Phys. Rev. Lett.}\ }\textbf {\bibinfo {volume} {82}},\ \bibinfo {pages} {5128} (\bibinfo {year} {1999})}\BibitemShut {NoStop}%
\bibitem [{\citenamefont {Madjarov}(2021)}]{madjarov}%
  \BibitemOpen
  \bibfield  {author} {\bibinfo {author} {\bibfnamefont {I.}~\bibnamefont {Madjarov}},\ }\href@noop {} {\enquote {\bibinfo {title} {Entangling, controlling, and detecting individual strontium atoms in optical tweezer arrays},}\ } (\bibinfo {year} {2021})\BibitemShut {NoStop}%
\bibitem [{\citenamefont {Wu}\ \emph {et~al.}(2022)\citenamefont {Wu}, \citenamefont {Kolkowitz}, \citenamefont {Puri},\ and\ \citenamefont {Thompson}}]{conversion}%
  \BibitemOpen
  \bibfield  {author} {\bibinfo {author} {\bibfnamefont {Y.}~\bibnamefont {Wu}}, \bibinfo {author} {\bibfnamefont {S.}~\bibnamefont {Kolkowitz}}, \bibinfo {author} {\bibfnamefont {S.}~\bibnamefont {Puri}}, \ and\ \bibinfo {author} {\bibfnamefont {J.~D.}\ \bibnamefont {Thompson}},\ }\href {\doibase 10.1038/s41467-022-32094-6} {\bibfield  {journal} {\bibinfo  {journal} {Nature Communications}\ }\textbf {\bibinfo {volume} {13}} (\bibinfo {year} {2022}),\ 10.1038/s41467-022-32094-6}\BibitemShut {NoStop}%
\bibitem [{\citenamefont {Scholl}\ \emph {et~al.}(2023{\natexlab{b}})\citenamefont {Scholl}, \citenamefont {Shaw}, \citenamefont {Tsai}, \citenamefont {Finkelstein}, \citenamefont {Choi},\ and\ \citenamefont {Endres}}]{conversion2}%
  \BibitemOpen
  \bibfield  {author} {\bibinfo {author} {\bibfnamefont {P.}~\bibnamefont {Scholl}}, \bibinfo {author} {\bibfnamefont {A.~L.}\ \bibnamefont {Shaw}}, \bibinfo {author} {\bibfnamefont {R.~B.-S.}\ \bibnamefont {Tsai}}, \bibinfo {author} {\bibfnamefont {R.}~\bibnamefont {Finkelstein}}, \bibinfo {author} {\bibfnamefont {J.}~\bibnamefont {Choi}}, \ and\ \bibinfo {author} {\bibfnamefont {M.}~\bibnamefont {Endres}},\ }\href@noop {} {\bibfield  {journal} {\bibinfo  {journal} {Nature}\ }\textbf {\bibinfo {volume} {622}},\ \bibinfo {pages} {273} (\bibinfo {year} {2023}{\natexlab{b}})}\BibitemShut {NoStop}%
\bibitem [{\citenamefont {de~Keijzer}\ \emph {et~al.}(2024)\citenamefont {de~Keijzer}, \citenamefont {Visser}, \citenamefont {Tse},\ and\ \citenamefont {Kokkelmans}}]{sse}%
  \BibitemOpen
  \bibfield  {author} {\bibinfo {author} {\bibfnamefont {R.}~\bibnamefont {de~Keijzer}}, \bibinfo {author} {\bibfnamefont {L.}~\bibnamefont {Visser}}, \bibinfo {author} {\bibfnamefont {O.}~\bibnamefont {Tse}}, \ and\ \bibinfo {author} {\bibfnamefont {S.}~\bibnamefont {Kokkelmans}},\ }\href@noop {} {\bibfield  {journal} {\bibinfo  {journal} {arXiv preprint arXiv:2401.11758}\ } (\bibinfo {year} {2024})}\BibitemShut {NoStop}%
\bibitem [{\citenamefont {Johansson}\ \emph {et~al.}(2012)\citenamefont {Johansson}, \citenamefont {Nation},\ and\ \citenamefont {Nori}}]{qutip}%
  \BibitemOpen
  \bibfield  {author} {\bibinfo {author} {\bibfnamefont {J.~R.}\ \bibnamefont {Johansson}}, \bibinfo {author} {\bibfnamefont {P.~D.}\ \bibnamefont {Nation}}, \ and\ \bibinfo {author} {\bibfnamefont {F.}~\bibnamefont {Nori}},\ }\href@noop {} {\bibfield  {journal} {\bibinfo  {journal} {Computer Physics Communications}\ }\textbf {\bibinfo {volume} {183}},\ \bibinfo {pages} {1760} (\bibinfo {year} {2012})}\BibitemShut {NoStop}%
\bibitem [{\citenamefont {Paszke}\ \emph {et~al.}(2019)\citenamefont {Paszke}, \citenamefont {Gross}, \citenamefont {Massa}, \citenamefont {Lerer}, \citenamefont {Bradbury}, \citenamefont {Chanan}, \citenamefont {Killeen}, \citenamefont {Lin}, \citenamefont {Gimelshein}, \citenamefont {Antiga}, \citenamefont {Desmaison}, \citenamefont {Kopf}, \citenamefont {Yang}, \citenamefont {DeVito}, \citenamefont {Raison}, \citenamefont {Tejani}, \citenamefont {Chilamkurthy}, \citenamefont {Steiner}, \citenamefont {Fang}, \citenamefont {Bai},\ and\ \citenamefont {Chintala}}]{torch}%
  \BibitemOpen
  \bibfield  {author} {\bibinfo {author} {\bibfnamefont {A.}~\bibnamefont {Paszke}}, \bibinfo {author} {\bibfnamefont {S.}~\bibnamefont {Gross}}, \bibinfo {author} {\bibfnamefont {F.}~\bibnamefont {Massa}}, \bibinfo {author} {\bibfnamefont {A.}~\bibnamefont {Lerer}}, \bibinfo {author} {\bibfnamefont {J.}~\bibnamefont {Bradbury}}, \bibinfo {author} {\bibfnamefont {G.}~\bibnamefont {Chanan}}, \bibinfo {author} {\bibfnamefont {T.}~\bibnamefont {Killeen}}, \bibinfo {author} {\bibfnamefont {Z.}~\bibnamefont {Lin}}, \bibinfo {author} {\bibfnamefont {N.}~\bibnamefont {Gimelshein}}, \bibinfo {author} {\bibfnamefont {L.}~\bibnamefont {Antiga}}, \bibinfo {author} {\bibfnamefont {A.}~\bibnamefont {Desmaison}}, \bibinfo {author} {\bibfnamefont {A.}~\bibnamefont {Kopf}}, \bibinfo {author} {\bibfnamefont {E.}~\bibnamefont {Yang}}, \bibinfo {author} {\bibfnamefont {Z.}~\bibnamefont {DeVito}}, \bibinfo {author} {\bibfnamefont {M.}~\bibnamefont {Raison}}, \bibinfo {author} {\bibfnamefont {A.}~\bibnamefont {Tejani}}, \bibinfo
  {author} {\bibfnamefont {S.}~\bibnamefont {Chilamkurthy}}, \bibinfo {author} {\bibfnamefont {B.}~\bibnamefont {Steiner}}, \bibinfo {author} {\bibfnamefont {L.}~\bibnamefont {Fang}}, \bibinfo {author} {\bibfnamefont {J.}~\bibnamefont {Bai}}, \ and\ \bibinfo {author} {\bibfnamefont {S.}~\bibnamefont {Chintala}},\ }in\ \href {http://papers.neurips.cc/paper/9015-pytorch-an-imperative-style-high-performance-deep-learning-library.pdf} {\emph {\bibinfo {booktitle} {Advances in Neural Information Processing Systems 32}}}\ (\bibinfo  {publisher} {Curran Associates, Inc.},\ \bibinfo {year} {2019})\ p.\ \bibinfo {pages} {8024}\BibitemShut {NoStop}%
\bibitem [{\citenamefont {Higgott}(2021)}]{pymatching}%
  \BibitemOpen
  \bibfield  {author} {\bibinfo {author} {\bibfnamefont {O.}~\bibnamefont {Higgott}},\ }\href@noop {} {\enquote {\bibinfo {title} {Pymatching: A python package for decoding quantum codes with minimum-weight perfect matching},}\ } (\bibinfo {year} {2021}),\ \Eprint {http://arxiv.org/abs/2105.13082} {arXiv:2105.13082 [quant-ph]} \BibitemShut {NoStop}%
\bibitem [{\citenamefont {Bombin}\ \emph {et~al.}(2012)\citenamefont {Bombin}, \citenamefont {Andrist}, \citenamefont {Ohzeki}, \citenamefont {Katzgraber},\ and\ \citenamefont {Martin-Delgado}}]{strongresilience}%
  \BibitemOpen
  \bibfield  {author} {\bibinfo {author} {\bibfnamefont {H.}~\bibnamefont {Bombin}}, \bibinfo {author} {\bibfnamefont {R.~S.}\ \bibnamefont {Andrist}}, \bibinfo {author} {\bibfnamefont {M.}~\bibnamefont {Ohzeki}}, \bibinfo {author} {\bibfnamefont {H.~G.}\ \bibnamefont {Katzgraber}}, \ and\ \bibinfo {author} {\bibfnamefont {M.~A.}\ \bibnamefont {Martin-Delgado}},\ }\href {\doibase 10.1103/PhysRevX.2.021004} {\bibfield  {journal} {\bibinfo  {journal} {Phys. Rev. X}\ }\textbf {\bibinfo {volume} {2}},\ \bibinfo {pages} {021004} (\bibinfo {year} {2012})}\BibitemShut {NoStop}%
\end{thebibliography}%

\end{document}